\documentclass[a4paper]{article}

\usepackage[english]{babel}
\usepackage[utf8x]{inputenc}
\usepackage[T1]{fontenc}

\usepackage[a4paper,top=3cm,bottom=2cm,left=3cm,right=3cm,marginparwidth=1.75cm]{geometry}

\usepackage{cite}
\usepackage{amsmath}
\usepackage{amssymb}
\usepackage{graphicx}
\usepackage{subcaption}
\usepackage[colorinlistoftodos]{todonotes}
\usepackage[colorlinks=true, allcolors=blue]{hyperref}
\usepackage[title]{appendix}
\usepackage{etoolbox}

\begin{document}
\title{Exact Decoherence Brought by One Internal Degree of Freedom: von Neumann Equation Approach and Examples}
\author{John Paul A. Besagas \thanks{jabesagas@up.edu.ph} , Jan Carlo L. Lima, Eric A. Galapon\thanks{eagalapon@up.edu.ph}\\Theoretical Physics Group, National Institute of Physics\\ University of the Philippines Diliman, Quezon City 1101, Philippines}

\maketitle

\begin{abstract}
In a quantum measurement setting, it is known that environment-induced decoherence theory describes the emergence of effectively classical 
features of the quantum system-measuring apparatus composite system when the apparatus is allowed to interact with the environment. In [E.A. Galapon {\it EPL} {\bf 113} 60007 (2016)], a measurement model is found to have the feature of inducing exact decoherence at a finite time via one internal degree of freedom of the apparatus provided that the apparatus is decomposed into a pointer and an inaccessible probe, with the pointer and the probe being in momentum-limited initial states. However, an issue can be raised against the model: while the factorization method of the time evolution operator used there is formally correct, it is not completely rigorous due to some unstated conditions on the validity of the factorization in the Hilbert space of the model. Furthermore, no examples were presented there in implementing the measurement scheme in specific quantum systems. The goal of this paper is to re-examine the model and confirm its features independently by solving the von Neumann equation for the joint state of the composite system as a function of time. This approach reproduces the joint state obtained in the original work, leading to the same conditions for exact decoherence and orthogonal pointer states when the required initial conditions on the probe and pointer are imposed. We illustrate the exact decoherence process in the measurement of observables of a spin-1/2 particle and a quantum harmonic oscillator by using the model.

\begin{description}
\item[PACS numbers:] 03.65.Ta, 03.65.Yz
\end{description}
\end{abstract}

\section{\label{sec:level1}Introduction}
The concept of measurement is important to us since it is our way to learn about nature. In the usual sense, a proper instrument registers a specific value of the property of an object under study. However, a different scenario arises in the measurement process performed on quantum systems. In the standard measurement scheme formulated by von Neumann, the Schr\"{o}dinger equation predicts a situation where the joint state of the quantum system of interest and the measuring apparatus becomes entangled at a later time. This situation presents a problem: while the entanglement between the system and apparatus implies that a correlation between the system and apparatus states has been established, the apparatus is however not in a state where it registers a definite read-out of the observable of the system. Furthermore, the problem is made worst by the inability of the Schr\"{o}dinger equation to describe how a definite outcome arises from the entangled state of the system-apparatus combined system \cite{zurek1, zurek2, zurek3, schlosshauer1, schlosshauer2, galapon1}. Such a conundrum is called the measurement problem; it is a persistent problem in quantum mechanics that has resulted to the different interpretations of the theory \cite{clarke, schlosshauer1, schlosshauer2}.   

In order to solve the measurement problem, one has to find a mechanism that accounts for the collapse of the entangled state of the system-apparatus combined system into a state that corresponds to a definite measurement outcome. The collapse of the quantum state can be classified into two kinds \cite{pessoa, omnes, namiki}. One is the collapse of the first kind, which pertains to the reduction of the state from quantum superposition into one of the states in the superposition. The other classification is called the collapse of the second kind or the statistical collapse of the quantum state, which pertains to the reduction of the state from quantum superposition into a statistical mixture. While the solution to the measurement problem demands a process that describes the collapse of the first kind, standard quantum mechanics however addresses the measurement problem through the collapse of the second kind, and it is known that the statistical collapse of the quantum state is described by the process of decoherence. A formalism of standard quantum mechanics that addresses the measurement problem by the concept of decoherence is called environment-induced decoherence theory (EIDT) \cite{zurek1, zurek2, zurek3, schlosshauer1, schlosshauer2, galapon1, clarke, pessoa, omnes, namiki, wallace, ballentine, tanona, unruh and zurek, venugopalan, connell, ford, myatt, abdul-redah, marques, arenz, sun, besagas}. Given the combined quantum system and apparatus in a correlated entangled state, EIDT claims that decoherence occurs by allowing the apparatus to interact with the large number of degrees of freedom of an environment; this leaves the combined system-apparatus in a statistical mixture which is defined by the states of the system and apparatus that, in the context of EIDT, are interpreted as effectively classical. Despite of its efforts in addressing the measurement problem, issues are being raised against EIDT due to the features of its models. In particular, (i) the suppression of the relevant coherences is only asymptotic (i.e. not exact) so that the emergent classicality from quantum measurement is only approximate, (ii) the pointer states are approximately orthogonal, which still implies ambiguity in the set of measurement outcomes \cite{galapon1,wallace}, and (iii) EIDT does not incorporate closed system-apparatus measurement models since the role of an external environment is necessary to induce decoherence \cite{galapon1,pessoa,ballentine,tanona}. With the ideas and criticisms of the theory, there is no general agreement among the scientific community on whether the measurement problem has been solved by EIDT \cite{schlosshauer1}. Several EIDT models have been considered in References \cite{zurek2, zurek3, unruh and zurek, venugopalan, connell, ford} while some experimental works involving decoherence in open quantum systems are discussed in References \cite{myatt,abdul-redah, marques}.

The criticisms (i)-(iii) of EIDT is addressed in a measurement model introduced in Reference \cite{galapon1}. The model assumed a system-apparatus measurement setting but the apparatus was designed in such a way that it is broken down into two separate subsystems: a pointer which gives the reading of the measured observable of the system, and a probe with a degree of freedom that is not observed in the entire measurement process. It was found in the measurement model that "exact decoherence" is induced at the level of system and pointer when the probe is initially in a momentum-limited state. There, exact decoherence means that the coherences between the states of the system and pointer are simultaneously and identically zero at a finite time.  Also, it was found that the pointer states become exactly orthogonal at a longer but finite time under the same initial state imposed on the pointer. In this model, the "exact" statistical collapse of the state of the system and pointer is achieved through one internal degree of freedom and then followed by the emergence of the set of unambiguous measurement outcomes. 

Inducing decoherence by one degree of freedom is already established in the literature. In fact, there are models at which decoherence can be induced via one degree of freedom external to the system of interest but the way how decoherence occurs in these models is only asymptotic. We mention references \cite{arenz, sun} for examples of such decoherence models. What makes the measurement model of \cite{galapon1} novel is the existence of a condition for inducing exact decoherence at a finite time, and the key for exact decoherence to occur is on the proper initial state of the probe. There, the probe assumes the role of the environment as a sink; as the interaction progresses, the information corresponding to the coherences of the reduced density matrix of the quantum system and apparatus pointer leaks out into the probe. Provided that the probe is initially in a momentum-limited state, all of the information corresponding to the coherences is secured in the probe at decoherence time. From the dynamics of the model, it is found that the probe becomes disconnected at decoherence time. As a result, the revival of coherences or recurrence does not happen in the model. Such disconnection process is a characteristic of standard measurement models; that is the apparatus is disconnected from the quantum system of interest at the end of measurement. While there exists a condition for exact decoherence to occur in the system and pointer of the measurement model of \cite{galapon1}, it is also found that this model exhibits asymptotic decoherence when the probe is initially prepared in a gaussian wavepacket in position space. There, the manner at which decoherence occurs (whether exact or asymptotic) is dictated by the initial state being imposed on the probe.

While the criticisms (i)-(iii) of EIDT is solved in the measurement model of \cite{galapon1}, there are issues that can be raised in this model. In particular, one may raise a question on the method of unravelling the joint state of the composite system as a function of measurement time. There, the time-evolution operator is factored by using the Zassenhaus formula \cite{casas}. While the factorization is formally correct, one needs to do more when rigor is insisted in applying the Zassenhaus formula in the Hilbert space of the model. However, such a task is non-trivial, and may, in fact, be intractable since the operators involved are unbounded. Another issue that can be raised in  \cite{galapon1} is the lack of examples that demonstrate the exact decoherence process and its effects in measuring observables of specific quantum systems. It is the goal of this paper to address these two issues. In order to do so, we consider the unitary quantum dynamics of the model by the approach of solving the von Neumann equation for the joint state of the system-probe-pointer as a function of measurement time under the assumptions that the subsystems are initially uncorrelated and the measurement Hamiltonian dominates the free Hamiltonians of the system, probe and pointer in the entire measurement process. We then show that this approach reproduces the joint state obtained in the original work, leading to the same conditions for exact decoherence and orthogonal pointer states when the required initial conditions on the probe and pointer are imposed. Moreover, using the measurement scheme of \cite{galapon1}, we present examples that illustrate exact decoherence in measuring the observable of spin-1/2 particle and quantum harmonic oscillator. In each example, we assume specific momentum-limited initial states of the probe and pointer and obtain exact time-dependent closed-forms for the decoherence factors and for the functions that measure the distinguishability of pointer states. These functions provide a picture of the dynamics of decoherence on the measurement model given the assumed initial conditions on the probe and pointer. Also, we compute for the probability densities for the measurement outcomes in the momentum representation of the pointer evaluated at orthogonality time. 

The paper is organized as follows. In section \ref{sec:level2}, we give a review of the measurement model discussed in \cite{galapon1}. In section \ref{sec:level3}, we derive and solve von Neumann equation for the joint density matrix of the system, probe and pointer as a function of measurement interaction time. In section \ref{sec:level4}, we present explicit examples wherein the scheme is implemented for the measurements of the $z$-component of the spin observable of a spin-1/2 particle and the energy observable of the quantum harmonic oscillator.

\section{\label{sec:level2}Review: Exact Decoherence Brought by One Internal Degree of Freedom of the Apparatus}

\begin{figure}[t!]
\centering
\includegraphics[width= 3.5in]{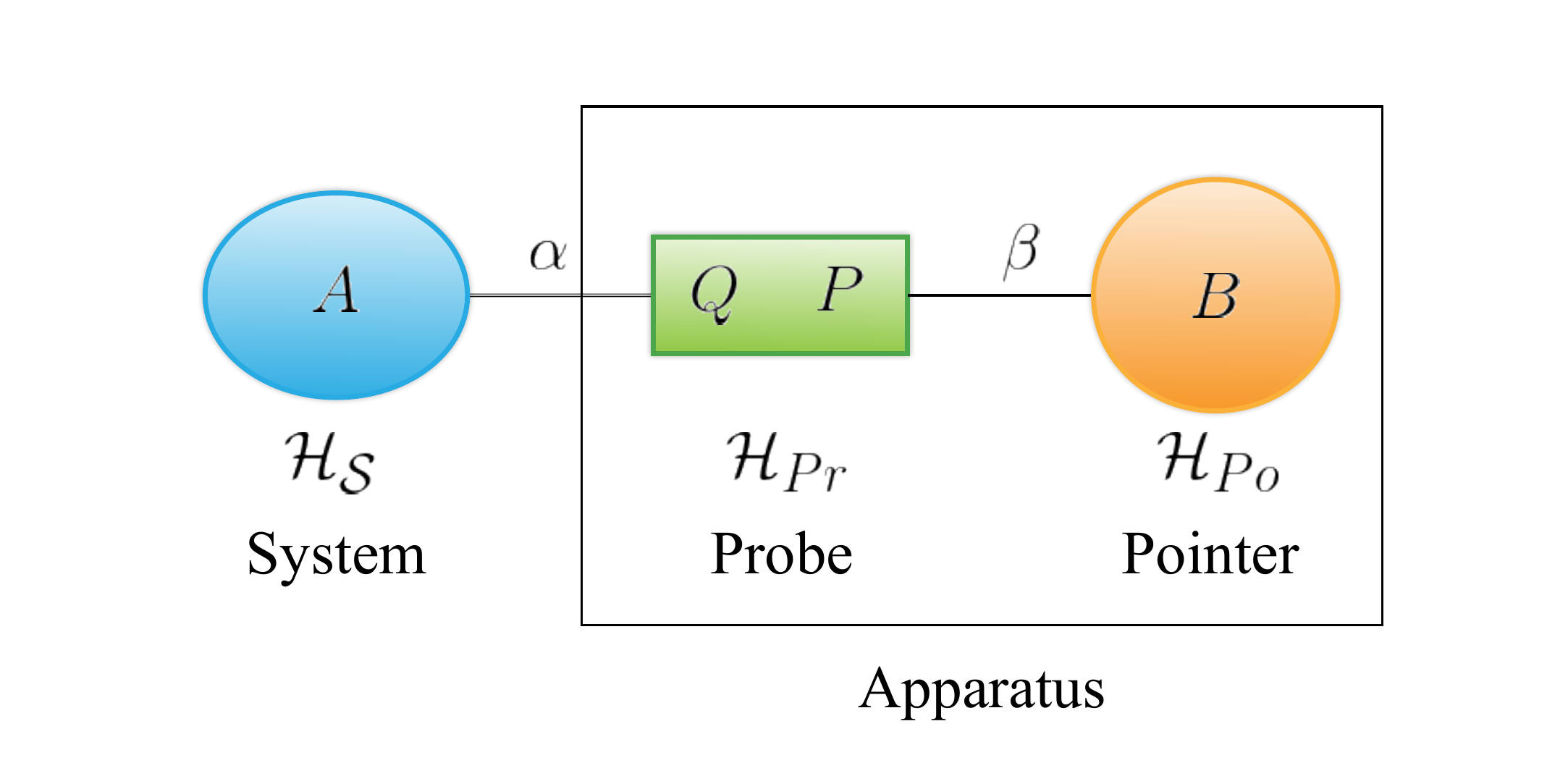}
\caption{Implementation of the measurement of a nondegenerate observable $A$ of a finite-dimensional quantum system $\mathcal{S}$ by means of a quantum apparatus that is decomposed into a probe and pointer.}\label{fig:model}
\end{figure}

We start by giving an overview of the measurement process that was introduced in \cite{galapon1}. A measurement  of a nondegenerate observable $A = \sum_k a_k |\varphi_k\rangle \langle \varphi_k|$ of a finite dimensional quantum system $\mathcal{S}$ is implemented by means of a measuring apparatus that is decomposed into a pointer and a probe. There, the pointer gives the read-out of the measured observable $A$, while the probe is the part of the apparatus with a degree of freedom that is unobservable in the entire process of the measurement. The system, probe and pointer form a closed composite system wherein these subsystems are organized as shown in Figure \ref{fig:model}. This composite system has a total Hilbert space $\mathcal{H} = \mathcal{H}_{\mathcal{S}} \otimes \mathcal{H}_{Pr} \otimes  \mathcal{H}_{Po}$, where $\mathcal{H}_{\mathcal{S}}$, $\mathcal{H}_{Pr}$ and $\mathcal{H}_{Po}$, are the respective Hilbert spaces of the system, probe and pointer. There, the Hilbert spaces $\mathcal{H}_{Pr}$ and $\mathcal{H}_{Po}$ are taken to be infinite-dimensional.

The measurement follows a von Neumann-like scheme \cite{busch} with measurement Hamiltonian given by
\begin{equation}\label{Hamiltonian}
H_M(t) = g(t) \,[\alpha\, A \otimes Q \otimes \mathbb{I}_{Po} + \beta \, \mathbb{I}_{\mathcal{S}} \otimes P \otimes B],
\end{equation}
where $Q$ and $P$ are the generalized position and momentum operators of the probe, $B$ is the pointer position observable, while $\alpha$ and $\beta$ are positive coupling constants. The time-dependent part $g(t)$ of the measurement Hamiltonian is taken to be localized in time; it is equal to $g_0>0$ at $t_{in} \leq t \leq t_f$, and is zero for other times. The composite system is prepared initially in a pure uncorrelated state $|\Psi_0\rangle = |\psi_{\mathcal{S}}\rangle \otimes |\psi_{Pr}\rangle \otimes |\Phi_{Po}\rangle$, where $|\psi_{\mathcal{S}}\rangle$, $|\psi_{Pr}\rangle$, and $|\Phi_{Po}\rangle$ are the respective initial states of the system, probe and pointer. It is assumed that $H_M(t)$ dominates the  free Hamiltonians of the system, probe and pointer during the duration of the measurement. Then the global state of the system, probe and pointer after the measurement is given by the joint density matrix
\begin{equation}\label{fin}
\rho_f = U\, |\Psi_0\rangle \langle \Psi_0|\,U^{\dagger},
\end{equation}
where 
\begin{equation}\label{tev1}
U=\exp\left(-\frac{i}{\hbar}\alpha g_0 \Delta\tau A\otimes Q \otimes \mathbb{I}_{Po} - \frac{i}{\hbar} \beta g_0 \Delta\tau \mathbb{I}_S \otimes P \otimes B\right),
\end{equation}
is the corresponding time-evolution operator. Since $U$ is an exponential of a sum of two non-commuting operators, it is factored out by using Zassenhaus formula \cite{casas}, which yields
\begin{eqnarray}\label{tev2}
U&=&\mathrm{e}^{\frac{i\alpha \beta g_0^2 \Delta \tau^2}{2\hbar} A\otimes \mathbb{I}_{Pr}\otimes B} 
\, \cdot \, \mathrm{e}^{-\frac{i}{\hbar} \alpha g_0 \Delta\tau A\otimes Q \otimes \mathbb{I}_{Po}}\cdot \mathrm{e}^{-\frac{i}{\hbar}\beta g_0\Delta\tau \mathbb{I}_S\otimes P \otimes B}.
\end{eqnarray}
Equation (\ref{tev2}) is the required form of the time-evolution operator to obtain the explicit form of $\rho_f$. By examining the terms in equation (\ref{tev2}), a coupling between the observables $A$ and $B$ of the system and pointer has emerged and it is indicated by the factor $\mathrm{e}^{\frac{i\alpha \beta g_0^2 \Delta \tau^2}{2\hbar} A\otimes \mathbb{I}_{Pr}\otimes B}$. In Reference \cite{galapon1}, the coupling between $A$ and $B$ is assumed to be independent of their respective coupling to the probe. This assumption is realized by choosing the coupling constant $\beta$ to be $\beta = 2\lambda/\alpha g_0^2 \Delta \tau^2$. There, $\lambda > 0$ is defined as the coupling constant between the observables $A$ and $B$ and is taken to be independent of $\alpha$. In getting the joint density matrix $\rho_f$, the position operators $Q$ and $B$ of the probe and the pointer are assumed to admit continuous spectrum in the entire real line and satisfy the eigenvalue relations $Q|q\rangle = q|q\rangle$ and $B|b\rangle = b|b\rangle$. Then, the joint density matrix $\rho_f$ as a function of measurement time $\Delta \tau$ can be written in the form

\begin{multline}\label{fin2}
\rho_f = \sum_{k,l} \langle \varphi_k|\psi_\mathcal{S}\rangle \langle \psi_\mathcal{S}|\varphi_l\rangle \,|\varphi_k\rangle \langle \varphi_l|\, \otimes\, \int_{-\infty}^{\infty} \mathrm{d}q \int_{-\infty}^{\infty} \mathrm{d}q' \,\mathrm{e}^{-\frac{i}{\hbar}\alpha g_0 \Delta \tau (a_k q - a_l q')}\, |q\rangle \langle q'|\\
\, \otimes\, \int_{-\infty}^{\infty} \mathrm{d}b \int_{-\infty}^{\infty} \mathrm{d}b'\, \mathrm{e}^{\frac{i\lambda}{\hbar} (a_k b - a_l b')} \langle b|\Phi_{Po}\rangle \langle \Phi_{Po}|b'\rangle\\
\times \langle q -2\lambda b/\alpha g_0 \Delta \tau|\psi_{Pr}\rangle \langle \psi_{Pr}|q' -2\lambda b'/\alpha g_0 \Delta \tau\rangle |b\rangle \langle b'|,
\end{multline}
by substituting equation (\ref{tev2}) to equation (\ref{fin}), inserting identity operators $\mathbb{I}_{\mathcal{S}} = \sum_k |\varphi_k\rangle \langle \varphi_k|$, $\mathbb{I}_{Pr} = \int_{-\infty}^{\infty} \mathrm{d}q\, |q\rangle \langle q|$, and $\mathbb{I}_{Po} = \int_{-\infty}^{\infty} \mathrm{d}b\, |b\rangle \langle b|$ and performing the necessary operations. 

In the measurement model of EIDT, the degrees of freedom of the environment are unobserved, so that the environment is averaged out from the correlated state of the system-apparatus-environment. This gives the reduced density matrix of the system and apparatus, which is necessary in analyzing how decoherence occurs \cite{zurek1, zurek2, zurek3, schlosshauer1, schlosshauer2, galapon1, clarke, pessoa, omnes, namiki, wallace, ballentine, tanona, unruh and zurek, venugopalan, connell, ford, myatt, abdul-redah, marques, arenz, sun}. In the measurement model of \cite{galapon1}, the degree of freedom of the probe is unobservable, so that it is traced out from the joint state $\rho_f$. This yields the reduced density matrix $\rho_{\mathcal{S}\otimes Po} = \mathrm{Tr}_{Pr} (\rho_f)$ of the system and pointer which has the form
\begin{equation} \label{spo}
\rho_{\mathcal{S} \otimes Po} = e^{\frac{i \lambda}{\hbar} A\otimes \mathbb{I}_{Pr}\otimes B} \, \rho_0^*\, e^{-\frac{i \lambda}{\hbar} A\otimes \mathbb{I}_{Pr}\otimes B},
\end{equation}
where
\begin{equation}
\rho_0^* = \sum_{k,l} \langle \varphi_k|\psi_{\mathcal{S}} \rangle \langle \psi_{\mathcal{S}} |\varphi_l \rangle\, |\varphi_k\rangle \langle \varphi_l| \, \otimes\, \int_{-\infty}^{\infty} \mathrm{d}b \int_{-\infty}^{\infty} \mathrm{d}b' \,\langle b|\Phi_{Po}\rangle\, \langle \Phi_{Po}|b'\rangle \,I_{kl}(b,b')\, |b\rangle \langle b'|.
\end{equation}
The functions $I_{kl}(b,b')$ are the decoherence factors and are generally given by
\begin{equation}\label{coherence}
I_{kl}(b,b') = \int_{-\infty}^{\infty} \mathrm{d}q  \,  \mathrm{e}^{-\frac{i}{\hbar} \alpha g_0 \Delta \tau (a_k - a_l)q}\\ \langle q - 2\lambda b/\alpha g_0 \Delta \tau|\psi_{Pr}\rangle\, \langle \psi_{Pr}|q - 2\lambda b'/\alpha g_0 \Delta \tau\rangle.
\end{equation}
The decoherence factors given by equation (\ref{coherence}) gives the measure of coherences of the system and pointer. Exact decoherence is induced to the system and pointer when $I_{kl}(b,b') = 0$ for all $k \neq l$. From now on, we denote the decoherence factors by $I_{k \neq l}(b,b')$.

The decoherence factors $I_{k \neq l}(b,b')$ is a Fourier integral of the initial state of the probe in $q$-space with the parameter $\alpha g_0 \Delta \tau$. As pointed out in \cite{galapon1}, the key for the vanishing of $I_{k \neq l}(b,b')$ is on the assumption that the initial state $\langle q|\psi_{Pr}\rangle$ of the probe is momentum-limited. By definition, a wavefunction $\langle q|\psi_{Pr}\rangle \in L^2(\mathbb{R})$ is momentum-limited if it can be represented by the following Fourier integral: $\langle q|\psi_{Pr}\rangle = \int_{-\hbar\kappa_0}^{\hbar \kappa_0} \mathrm{d}p\, \mathrm{e}^{ipq/\hbar}\, \langle p|\psi_{Pr}\rangle$, where $\langle p|\psi_{Pr}\rangle$ is the state of the probe in the momentum representation and $\kappa_0$ is a finite and positive real constant \cite{galapon1}. A theorem in the theory of entire functions states that the complex plane extension of the wavefunction $\langle q|\psi_{Pr}\rangle = \int_{-\hbar\kappa_0}^{\hbar \kappa_0} \mathrm{d}p\, \mathrm{e}^{ipq/\hbar}\, \langle p|\psi_{Pr}\rangle$ is entire and exponential of type $\kappa_0$ (See Reference \cite{boas} or Appendix for the Paley-Wiener theorem). Then the following Lemma is used to deduce the condition for the vanishing of $I_{k \neq l}(b,b')$:
\newtheorem{lem}{Lemma}[section]
\begin{lem}
\label{exptype}
	Let $f(z)$ be entire and exponential of type $\tau > 0$, and $\int_{-\infty}^{\infty} |f(x)|\, \mathrm{d}x = M < \infty$. Then $\int_{-\infty}^{\infty} \mathrm{e}^{iax} f(x)\, \mathrm{d}x = 0$, for all $|a|>\tau$.
\end{lem}
Under the assumption that the initial state of the probe is momentum-limited of type $\kappa_0$, then from \ref{exptype}, all $I_{k \neq l}(b,b')$ vanish simultaneously and identically for $\Delta \tau > \Delta \tau_D$, where $\Delta \tau_D$ is the decoherence time
\begin{equation}\label{deco}
\Delta \tau_D = \frac{2 \kappa_0 \hbar}{\alpha g_0 a_0}.
\end{equation}
There, the quantity $a_0$ is given by $a_0 = \min\{(a_k - a_l), a_k > a_l\}$. At decoherence time, the probe becomes disconnected from the system and pointer. Moreover, the reduced density matrix $\rho_0^*$ becomes separable and mixed. The system and pointer becomes uncorrelated at decoherence time. 

After exact decoherence has occurred, the final establishment of correlation between the preferred states of system and the pointer progresses and the full reduced density matrix of the system and pointer becomes
\begin{equation}
\rho_{\mathcal{S}\otimes Po} = \sum_{k} |\langle \varphi_k|\psi_{\mathcal{S}}\rangle|^2 \, |\varphi_k\rangle \langle \varphi_k| \otimes\, \rho_k 
\end{equation}
where $\rho_k$'s are the pointer states
\begin{equation}\label{pointers}
\rho_k = \int_{-\infty}^{\infty} \mathrm{d}b \int_{-\infty}^{\infty} \mathrm{d}b' \,\langle b|\Phi_{Po}\rangle\, \langle \Phi_{Po}|b'\rangle
I_{0}(b,b') \, \mathrm{e}^{\frac{i\lambda}{\hbar} a_k(b - b')}\, |b\rangle \langle b'|.
\end{equation}
Here, $I_0(b, b') = I_{kk}(b,b')$. In considering the mutual orthogonality of two distinct pointer states $\rho_k$ and $\rho_l$, their product $\rho_k\rho_l$ is investigated, which is shown to have the form
\begin{equation}
\rho_k \rho_l = \int_{-\infty}^{\infty} \mathrm{d}b \int_{-\infty}^{\infty} \mathrm{d}b' \, \mathrm{e}^{\frac{i\lambda}{\hbar}(a_k b - a_l b')}\, \langle b|\Phi_{Po}\rangle \langle \Phi_{Po}|b'\rangle\, S_{kl}(b,b')\ |b\rangle \langle b'|
\end{equation}
where the function $S_{kl}(b,b')$ is given by
\begin{equation}
S_{kl}(b,b') = \int_{-\infty}^{\infty} \mathrm{d}b'' \mathrm{e}^{-\frac{i\lambda}{\hbar}(a_k - a_l)b''} |\langle b''|\Phi_{Po}\rangle|^2 F(b-b'') F(b''-b').
\end{equation}
There, the functions $F(\eta)$ given by $F(\eta) = \int_{-\kappa_0\hbar}^{\kappa_0 \hbar} \mathrm{d}p\, \mathrm{e}^{2i\lambda \eta p/\alpha g_0 \Delta \tau \hbar}\, |\langle p|\psi_{Pr}\rangle|^2$ are assumed to be real-valued and even functions of $\eta$ for the sake of simplicity. The pointer states are mutually orthogonal when all $S_{k \neq l}(b,b') = 0$. Under the assumption that the initial state of the probe is momentum-limited of type $\kappa_0$, the complex plane extension of $F(\eta)$ is entire and exponential of type $2\lambda \kappa_0/\alpha g_0 \Delta \tau$. Under this condition, it is found in\cite{galapon1} that exact orthogonality of pointer states is achieved if the initial state of the pointer $\langle b|\Phi_{Po}\rangle$ is also momentum-limited such that its complex extension is entire and exponential of type $b_0>0$. With the application of Lemma \ref{exptype}, it is found that $S_{k \neq l}(b,b') = 0$ for $\Delta \tau > \Delta \tau_O$, where 
\begin{equation}\label{orth}
\Delta \tau_O = \frac{4 \kappa_0 \hbar}{\alpha g_0 (a_0 - 2 b_0\hbar/\lambda)},
\end{equation}
is the orthogonality time. Moreover, exact orthogonality of pointer states requires that the coupling constant $\lambda$ should satisfy 
\begin{equation}\label{orth2}
\lambda > \frac{2b_0\hbar}{a_0}
\end{equation}
By comparing the decoherence and orthogonality times, it is found that $\Delta \tau_O > \Delta \tau_D$. This means that the reduced density matrix of the system and pointer has exactly decohered at $\Delta \tau > \Delta \tau_D$. Then this is followed by the emergence of unambiguous outcomes at measurement times $\Delta \tau > \Delta \tau_O$. 

One may have an interest on the meaning of the orthogonality condition given by equation (\ref{orth2}) in the measurement of the observable $A$ of the quantum system $\mathcal{S}$ using the measurement scheme of \cite{galapon1}. In orthodox quantum measurement theory, it is known that a projective measurement of an observable of a quantum system can be realized through the output observable (or pointer) of the measuring apparatus when the pointer states corresponding to different outcomes of the system observable have distinct supports. This distinction between the pointer states follows when the minimum difference between the eigenvalues of the Hermitian operator representing the system observable is greater than the ratio of the initial uncertainty of the output observable of the apparatus to the coupling constant between the system and apparatus \cite{busch, kofman}. In the measurement scheme of \cite{galapon1}, the orthogonality condition given by equation (\ref{orth2}) implies the condition for the projective measurement of the observable of $\mathcal{S}$. There, the output observable of the pointer is its momentum, with an initial uncertainty equal to $2b_0 \hbar$ (the initial uncertainty in the pointer momentum follows from the assumption that the pointer is initially prepared in a momentum-limited state with type $b_0$). Hence, when condition (\ref{orth2}) is satisfied in the measurement scheme of \cite{galapon1}, a projective measurement of the observable of $\mathcal{S}$ can be realized through the apparatus pointer at measurement times greater than or equal to the orthogonality time. In the examples to be presented in later sections, we support this implication by showing that the momentum representation of different pointer states have distinct supports at orthogonality time. 

In the measurement scheme of \cite{galapon1}, the reduced density matrix $\rho_{\mathcal{S} \otimes Po}$ of the system and pointer is important since it is needed in analyzing how decoherence occurs on this compound subsystem. This reduced density matrix is obtained from the joint final state $\rho_f$ of the system-probe-pointer by tracing out the probe. In \cite{galapon1}, the joint state $\rho_f$ is obtained by factoring the time-evolution operator $U$ by Zassenhaus formula, and the resulting factorized form of $U$ is acted on the joint initial state of the system-probe-pointer. The factorization method via Zassenhaus formula is formally correct, however it is not rigorous because of the unstated conditions on the validity of the method in the Hilbert space of the measurement model. Hence, this may raise a question on the validity of the features of the measurement scheme of \cite{galapon1} (i.e. conditions for exact decoherence and exactly orthogonal pointer states). In the next section, we confirm the global state of the system, probe and pointer as a function of measurement time by solving the corresponding von Neumann equation given that the subsystems are initially uncorrelated.

\section{\label{sec:level3} On solving the von Neumann equation for the joint state of system, probe and pointer}

Now we re-examine the unitary dynamics of the system-probe-pointer measurement model by solving the von Neumann equation for the joint state of the system, probe and pointer as a function of measurement time. Since the system, probe and pointer comprise a closed composite quantum system and the self Hamiltonians of the subsystems are assumed to have negligible effect on the time-evolution of the state of the composite system, then the joint state $\rho_f = |\Psi(t)\rangle \langle \Psi(t)|$ is a solution to the von Neumann equation
\begin{equation}\label{vne}
i\hbar\, \frac{\partial \rho_f}{\partial t} = [H_M (t),\rho_f],
\end{equation}
where $H_M(t)$ is the measurement Hamiltonian given by equation (\ref{Hamiltonian}). 

Let $\{|\varphi_k\rangle\}$, $\{|q \rangle\}$ and $\{|b\rangle\}$ be orthonormal bases in the system, probe and pointer Hilbert spaces, respectively. Then we can express the final state $|\Psi(t)\rangle$ in the form
\begin{equation}\label{psi}
|\Psi(t)\rangle = \sum_k \int_{-\infty}^{\infty} \mathrm{d}q \int_{-\infty}^{\infty} \mathrm{d}b\, (|\varphi_k\rangle \otimes |q\rangle \otimes |b\rangle)\, f_k(q,b;t),
\end{equation}
where we define the wavefunction $f_k(q,b;t) = (\langle \varphi_k| \otimes \langle q| \otimes \langle b|) |\Psi(t)\rangle$.  From equation (\ref{psi}), the corresponding density matrix $\rho_f$ has the form
\begin{equation}\label{fdm}
\rho_f = \sum_{k,l} \int_{-\infty}^{\infty} \mathrm{d}q \int_{-\infty}^{\infty} \mathrm{d}q' \int_{-\infty}^{\infty} \mathrm{d}b \int_{-\infty}^{\infty} \mathrm{d}b'\, (|\varphi_k\rangle \langle \varphi_l|\, \otimes\, |q\rangle \langle q'|\, \otimes\, |b\rangle \langle b'|)\, f_k(q,b;t)\, f_l^*(q',b';t). 
\end{equation}
This form of $\rho_f$ contains off-diagonal and diagonal terms, which we denote by $\rho_{kl}(q,q',b,b';t)$ and $\rho_{kk}(q,b;t)$ respectively. Given equation (\ref{fdm}), it can be shown that these terms take the form
\begin{equation}\label{offdiag}
\rho_{kl}(q,q',b,b';t) = f_k(q,b;t)\, f_l^*(q',b';t),
\end{equation}
and 
\begin{equation}\label{diag}
\rho_{kk}(q,b;t) = C_{kk}(q,b;t),
\end{equation}
where $C_{kk}(q,b;t) = |f_k(q,b;t)|^2$. 

Here, our goal is to obtain $\rho_{kl}(q,q',b,b';t)$ and $\rho_{kk}(q,b;t)$ from their respective equations of motion that can be derived from the von Neumann equation (\ref{vne}). This is done by substituting $H_M(t)$ and equation (\ref{fdm}) to equation (\ref{vne}) and tracing out the probe and then the pointer on both sides of the von Neumann equation. With lengthy but straightforward steps, the trace operation on the left hand side of equation (\ref{vne}) yields
\begin{equation}\label{lhs}
i \hbar\, \mathrm{Tr}_{Po} \mathrm{Tr}_{Pr} \left(\frac{\partial \rho_f}{\partial t} \right) = i \hbar \,\sum_{k,l} |\varphi_k\rangle \langle \varphi_l|\,\otimes\, \int_{-\infty}^{\infty} \mathrm{d}b \int_{-\infty}^{\infty} \mathrm{d}q\,\, \frac{\partial C_{kl}(q,b;t)}{\partial t}.
\end{equation}
where we define $C_{kl}(q,b;t) =f_k(q,b;t)\, f_l^*(q,b;t)$. On the other hand, the trace operation on the right hand side of equation (\ref{vne}) results to the form of $\Lambda = \mathrm{Tr}_{Po} \mathrm{Tr}_{Pr} [H_M(t), \rho_f]$ given by
\begin{equation}\label{rhs}
\Lambda =  g(t) \sum_{k,l} |\varphi_k\rangle \langle \varphi_l| \int_{-\infty}^{\infty} \mathrm{d}q \int_{-\infty}^{\infty} \mathrm{d}b \, \left[ \alpha\, (a_k -a_l)\, q\, C_{kl}(q,b;t) + \frac{\hbar}{i}\, \beta\, b\, \,\frac{\partial C_{kl}(q,b;t)}{\partial q}\right]
\end{equation}
Equating (\ref{lhs}) and (\ref{rhs}) results to the partial differential equation for $C_{kl}(q,b;t)$
\begin{equation}\label{vneqb}
\alpha \, (a_k - a_l)\, g(t)\, q\, C_{kl}(q,b;t)\\ +\, \frac{\hbar}{i}\beta\, g(t)\, b\, \frac{\partial C_{kl}(q,b;t)}{\partial q} = i\hbar\, \frac{\partial C_{kl}(q,b;t)}{\partial t},
\end{equation}
for $k \neq l$. Equation (\ref{vneqb}) is a partial differential equation that is linear and first order in probe position variable $q$ and time $t$. Then $\rho_{kl}(q,q',b,b';t)$ can be obtained by solving equation (\ref{vneqb}) given the initial condition
\begin{equation}\label{inc}
C_{kl}(q,b;0) = \langle \varphi_k|\psi_{\mathcal{S}} \rangle \langle \psi_{\mathcal{S}} |\varphi_l \rangle \, |\langle q|\psi_{Pr}\rangle|^2 \,|\langle b|\Phi_{Po}\rangle|^2.
\end{equation}
With the use of method of characteristics \cite{adzievski} (see Appendix for the steps), it can be shown that 
\begin{equation}\label{soln}
C_{kl}(q,b;t) =  \langle \varphi_k|\psi_{\mathcal{S}} \rangle \langle \psi_{\mathcal{S}} |\varphi_l \rangle \, |\langle q - \beta b g_0 t|\psi_{Pr}\rangle|^2  |\langle b|\Phi_{Po}\rangle|^2 \, \mathrm{e}^{-\frac{i}{\hbar} \alpha g_0 t (a_k - a_l)q}\,
\mathrm{e}^{\frac{i}{2\hbar} \alpha \beta g_0^2 t^2 (a_k - a_l)b}.
\end{equation}
Moreover, it follows from the definition of $C_{kl}(q,b;t)$ that 
\begin{equation}\label{soln2}
f_k(q,b;t) = \langle \varphi_k|\psi_{\mathcal{S}} \rangle \langle q - \beta b g_0 t|\psi_{Pr}\rangle \langle b|\Phi_{Po}\rangle \mathrm{e}^{-\frac{i}{\hbar} \alpha g_0 t a_k q}\, \mathrm{e}^{\frac{i}{2\hbar} \alpha \beta g_0^2 t^2 a_k b}.
\end{equation}
With these results, then the explicit expressions for the off-diagonal terms of $\rho_f$ are
\begin{multline}\label{od}
\rho_{kl}(q,q',b,b';t) = \langle \varphi_k|\psi_{\mathcal{S}} \rangle \, \langle \psi_{\mathcal{S}} |\varphi_l \rangle \,  \langle q - \beta b g_0 t|\psi_{Pr}\rangle \langle \psi_{Pr}|q' - \beta b' g_0 t\rangle\, \langle b|\Phi_{Po}\rangle\,  \langle \Phi_{Po}|b'\rangle\\
\times \mathrm{e}^{-\frac{i}{\hbar} \alpha g_0 t (a_k q - a_l q')}\,\mathrm{e}^{\frac{i}{2\hbar} \alpha \beta g_0^2 t^2 (a_k b - a_l b')}
\end{multline}
In a similar manner, $\rho_{kk}(q,b;t)$ can be obtained by solving the partial differential equation
\begin{equation}\label{vneqb2}
\frac{\hbar}{i}\beta\, g(t)\, b\, \frac{\partial C_{kk}(q,b;t)}{\partial q} = i\hbar\, \frac{\partial C_{kk}(q,b;t)}{\partial t}.
\end{equation}
subject to the initial condition given by equation (\ref{inc}) with $k=l$. It is straightforward to show that (see Appendix)
\begin{equation}\label{d}
\rho_{kk}(q,b;t) = |\langle \varphi_k|\psi_{\mathcal{S}} \rangle|^2 \, |\langle q - \beta b g_0 t|\psi_{Pr}\rangle|^2\, |\langle b|\Phi_{Po}\rangle|^2.
\end{equation}
With these results, then the explicit form of the joint state $\rho_f$ as a function of measurement time $t$ is given by
\begin{multline}
\rho_f = \sum_{k,l} \langle \varphi_k|\psi_\mathcal{S}\rangle \langle \psi_\mathcal{S}|\varphi_l\rangle \,|\varphi_k\rangle \langle \varphi_l| \,\otimes\, \int_{-\infty}^{\infty} \mathrm{d}q \int_{-\infty}^{\infty} \mathrm{d}q' \,\mathrm{e}^{-\frac{i}{\hbar}\alpha g_0 t (a_k q - a_l q')}\, |q\rangle \langle q'| \\ \otimes\, \int_{-\infty}^{\infty} \mathrm{d}b \int_{-\infty}^{\infty} \mathrm{d}b'\, \mathrm{e}^{\frac{i\lambda}{\hbar} (a_k b - a_l b')} \langle b|\Phi_{Po}\rangle \langle \Phi_{Po}|b'\rangle \langle q -2\lambda b/\alpha g_0 t|\psi_{Pr}\rangle \langle \psi_{Pr}|q' -2\lambda b'/\alpha g_0 t\rangle \, |b\rangle \langle b'|, 
\end{multline}
where we have set $\beta = 2\lambda / \alpha g_0^2 t^2$. This is the same as that of $\rho_f$ given by equation (\ref{fin2}), with $\Delta \tau = t$. Consequently, we can obtain the same form of the reduced density matrix of the system and pointer (equation \ref{spo}), decoherence factors (equation \ref{coherence}), and pointer states (equation \ref{pointers}). Then we can deduce the same conditions for exact decoherence and mutual orthogonality of pointer states by imposing the necessary assumptions on the initial states of the probe and pointer, that is, both of them should have momentum-limited initial states. 

It is also of interest to consider how decoherence takes place in the quantum system $\mathcal{S}$. This requires obtaining the reduced density matrix of $\mathcal{S}$ as a function of measurement interaction time. We denote it by $\rho_{\mathcal{S}}$. By tracing out the probe and pointer from $\rho_f$, we have
\begin{equation}\label{rhos}
\rho_{\mathcal{S}} = \sum_{k,l} |\varphi_k\rangle \langle \varphi_l|\, \langle \varphi_k|\psi_{\mathcal{S}} \rangle \langle \psi_{\mathcal{S}} |\varphi_l \rangle \int_{-\infty}^{\infty} \mathrm{d}b\, \mathrm{e}^{-\frac{i \lambda}{\hbar}(a_k - a_l)b}\,|\langle b|\Phi_{Po}\rangle|^2 \int_{-\infty}^{\infty} \mathrm{d}q\,\,  \mathrm{e}^{-\frac{i}{\hbar} \alpha g_0 t (a_k - a_l)q} |\langle q |\psi_{Pr}\rangle|^2 
\end{equation}
We consider the decoherence factors of $\rho_{\mathcal{S}}$. There are two possible initial conditions for the vanishing of the coherences of the system. One condition is when the initial state $\langle q|\psi_{Pr}\rangle$ of the probe is momentum-limited of type $\kappa_0$. By applying Lemma \ref{exptype}, the off-diagonal elements of $\rho_{\mathcal{S}}$ vanish identically for $\Delta \tau > \Delta \tau_D$, where $\Delta \tau_D$ is just the decoherence time given by equation (\ref{deco}). The other condition for exact decoherence to occur on system is when the initial state of the pointer $\langle b|\Phi_{Po}\rangle$ is also momentum-limited with type $b_0$. Then the off-diagonal terms of $\rho_{\mathcal{S}}$ vanish identically when the coupling constant $\lambda$ satisfies the condition $\lambda > 2b_0\hbar/a_0$. Thus, exact decoherence occurs at the level of the system $\mathcal{S}$ when either of the probe or the pointer is initially in a momentum-limited state. This feature of the measurement model has not been seen in \cite{galapon1}. 

\section{\label{sec:level4}Examples}
In this section, we illustrate the exact decoherence process in implementing the scheme of \cite{galapon1} in the measurement on spin-1/2 quantum particle and a quantum harmonic oscillator. 
\subsection{Spin-1/2 Quantum Particle}
\begin{figure}[t]
    \centering
    \begin{subfigure}[b]{0.48\textwidth}
        \includegraphics[width=\textwidth]{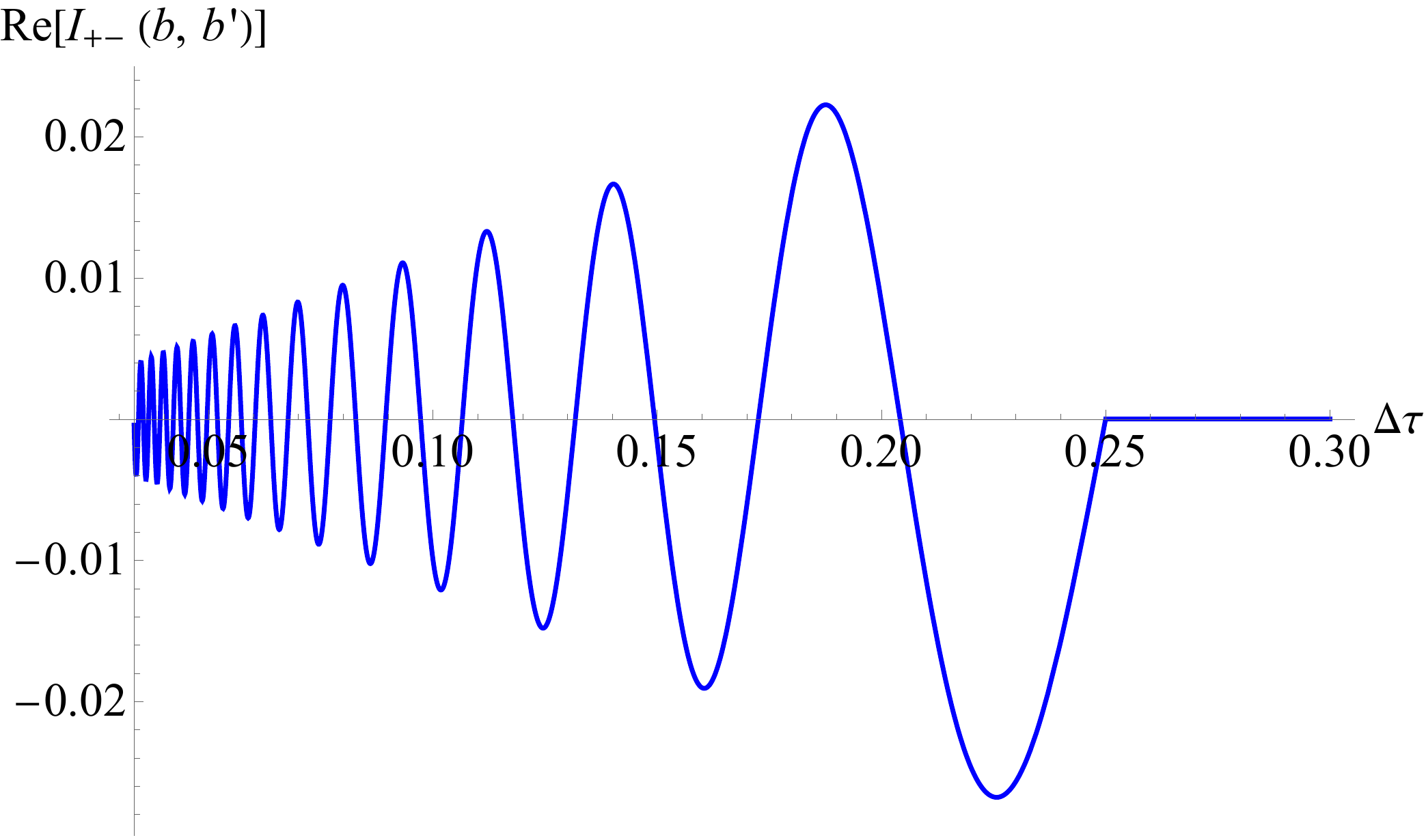}
        \caption{}
        \label{fig:CoherencePMreal}
    \end{subfigure}
    ~ 
    \begin{subfigure}[b]{0.48\textwidth}
        \includegraphics[width=\textwidth]{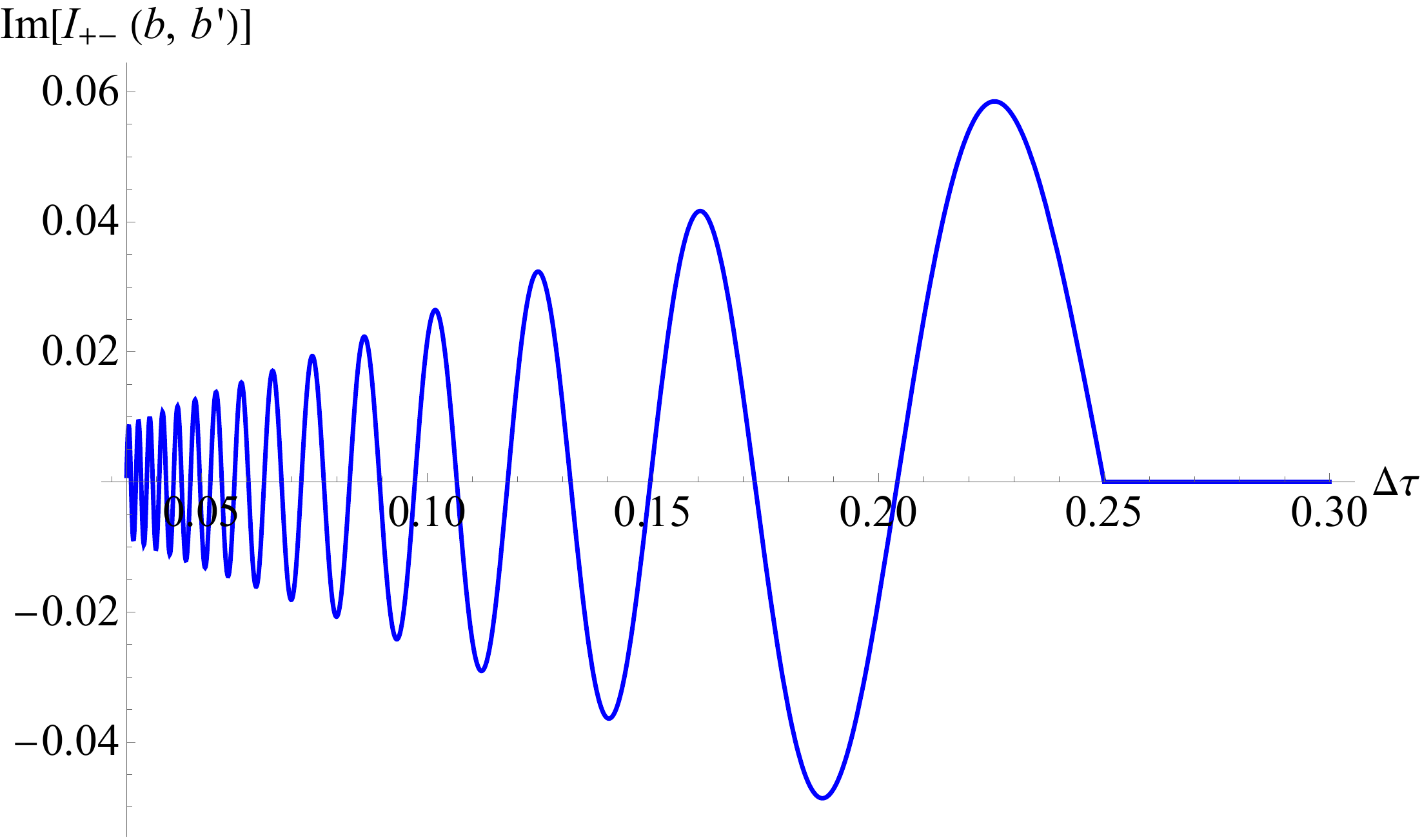}
        \caption{}
        \label{fig:CoherencePMIm}
    \end{subfigure}
    \caption{Plot of (a) real part, and (b) imaginary part of the decoherence factor $I_{+-}(b,b')$ given by equation (\ref{qubcohexact}) as a function of interaction time $\Delta \tau$ for the measurement of the observable $S_z$ of a qubit, with parameter values $\lambda = 4$ $\alpha = 1$, $\kappa_0 = 0.25$, $g_0 = 2$, $b = 1.5$, and $b' = -2$. Both real and imaginary parts of $I_{+-}(b,b')$ vanish at and beyond $\Delta \tau = 0.25$. Evaluation of the $I_{+-}(b,b')$ by quadratures leads to the same plots as above.}\label{fig:qubitdeco}
\end{figure}

We consider the implementation of the scheme in measuring the $z$-component of spin $S_z$ of a spin-1/2 quantum particle. Recall that $S_z$ is non-degenerate, and has eigenstates and corresponding eigenvalues defined by the relation $ S_z|\pm\rangle = \pm \frac{\hbar}{2} |\pm \rangle$ \cite{griffiths}. The difference between the eigenvalues of $S_z$ is just $\hbar$. We let the probe in the initial state
\begin{equation}\label{ipr}
\langle q| \psi_{Pr} \rangle = \frac{1}{\sqrt{\kappa_0 \pi}} \frac{\sin(\kappa_0 q)}{q}, 
\end{equation}
where $\kappa_0 > 0$. Moreover, we let the initial state of the pointer to be
\begin{equation}\label{ipo}
\langle b| \Phi_{Po} \rangle = \frac{1}{\sqrt{b_0 \pi}} \frac{\sin(b_0 b)}{b}, 
\end{equation}
where $b_0>0$. The wavefunctions given by equations (\ref{ipr}) and (\ref{ipo}) have momentum representations that have compact supports and therefore, have complex plane extensions that are entire and exponential of type $\kappa_0$ and $b_0$ respectively (see Appendix). Then from equation (\ref{coherence}), the corresponding decoherence factors of reduced density matrix of the system-pointer are given by
\begin{equation}\label{coh1}
I_{+-}(b,b') =  \frac{1}{\kappa_0 \pi} \int_{-\infty}^{\infty} \mathrm{d}q \, \, \, \mathrm{e}^{-i \alpha g_0 \Delta \tau q} \frac{\sin \left[\kappa_0(q -\frac{2 \lambda b}{\alpha g_0 \Delta \tau})\right]}{(q - \frac{2 \lambda b}{\alpha g_0 \Delta \tau})} \frac{\sin \left[\kappa_0(q -\frac{2 \lambda b'}{\alpha g_0 \Delta \tau})\right]}{(q - \frac{2 \lambda b'}{\alpha g_0 \Delta \tau})},
\end{equation}
and $I_{-+}(b,b') = I_{+-}^*(b,b')$. From \cite{galapon1} (or by using Lemma \ref{exptype}), it can be argued that $I_{+-}(b,b')$ and $I_{-+}(b,b')$ are both exactly zero for measurement times $\Delta \tau > 2\kappa_0/\alpha g_0$, where the minimum $\Delta \tau_D = 2\kappa_0/\alpha g_0$ is the corresponding decoherence time. However, we show here that $I_{+-}(b,b')$ and $I_{-+}(b,b')$ vanish for times $\Delta \tau \geq \Delta \tau_D$ instead of the former condition $\Delta \tau > \Delta \tau_D$ by evaluating their closed forms. Using the techniques discussed in \cite{galapon2, lima}, we have
\begin{equation}\label{qubcohexact}
I_{+-}(b,b') = \frac{i\alpha g_0 \Delta \tau}{4 \lambda \kappa_0}\, U(b,b')\, \Theta(2\kappa_0 - \alpha g_0 \Delta \tau),
\end{equation}
where
\begin{equation}
U(b,b') = \frac{\left[\,\mathrm{e}^{-2i\lambda \left(\frac{\kappa_0}{\alpha g_0 \Delta \tau}(b-b') + b' \right)} - \mathrm{e}^{2i\lambda \left(\frac{\kappa_0}{\alpha g_0 \Delta \tau}(b-b') - b \right)}\right]}{b-b'},
\end{equation}
and $\Theta(x)$ is the Heaviside step function. The presence of $\Theta(2\kappa_0 - \alpha g_0 \Delta \tau) $ verifies the condition for the vanishing of $I_{+-}(b,b')$ and $I_{-+}(b,b')$ as a consequence of Lemma \ref{exptype}. However, if we set $\Delta \tau = 2\kappa_0/\alpha g_0$, we find that $U(b,b') = 0$ so that both $I_{+-}(b,b')$ and $I_{-+}(b,b')$ are equal to zero.  Thus, the relevant coherences vanish for $\Delta \tau \geq 2\kappa_0/\alpha g_0$. Figure \ref{fig:qubitdeco} shows the plot of the real and imaginary parts of $I_{+-}(b,b')$ as functions of measurement time $\Delta \tau$. Both real and imaginary parts of $I_{+-}(b,b')$ oscillate with increasing amplitude in time but eventually vanish for $\Delta \tau \geq 0.25$, with the minimum exactly equal to the corresponding value of the decoherence time (equation \ref{deco}) given the assumed values of the relevant parameters. A quadrature evaluation of equation (\ref{coh1}) can be implemented and yields results that are exactly the same as that of the plots in Figure \ref{fig:qubitdeco}. 

\begin{figure}[t]
    \centering
    \begin{subfigure}[b]{0.48\textwidth}
        \includegraphics[width=\textwidth]{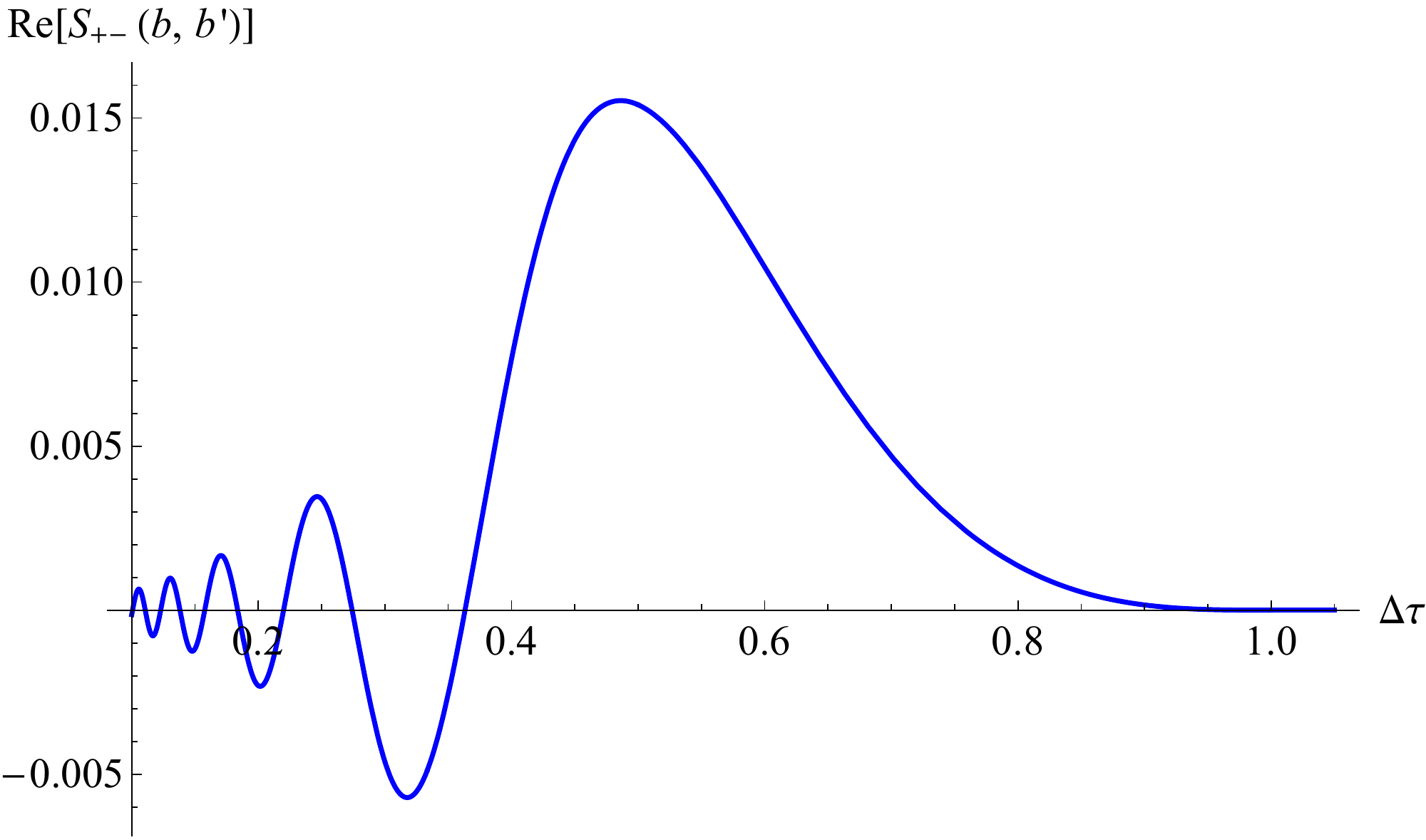}
        \caption{}
        \label{fig:orthoreal}
    \end{subfigure}
    ~ 
    \begin{subfigure}[b]{0.48\textwidth}
        \includegraphics[width=\textwidth]{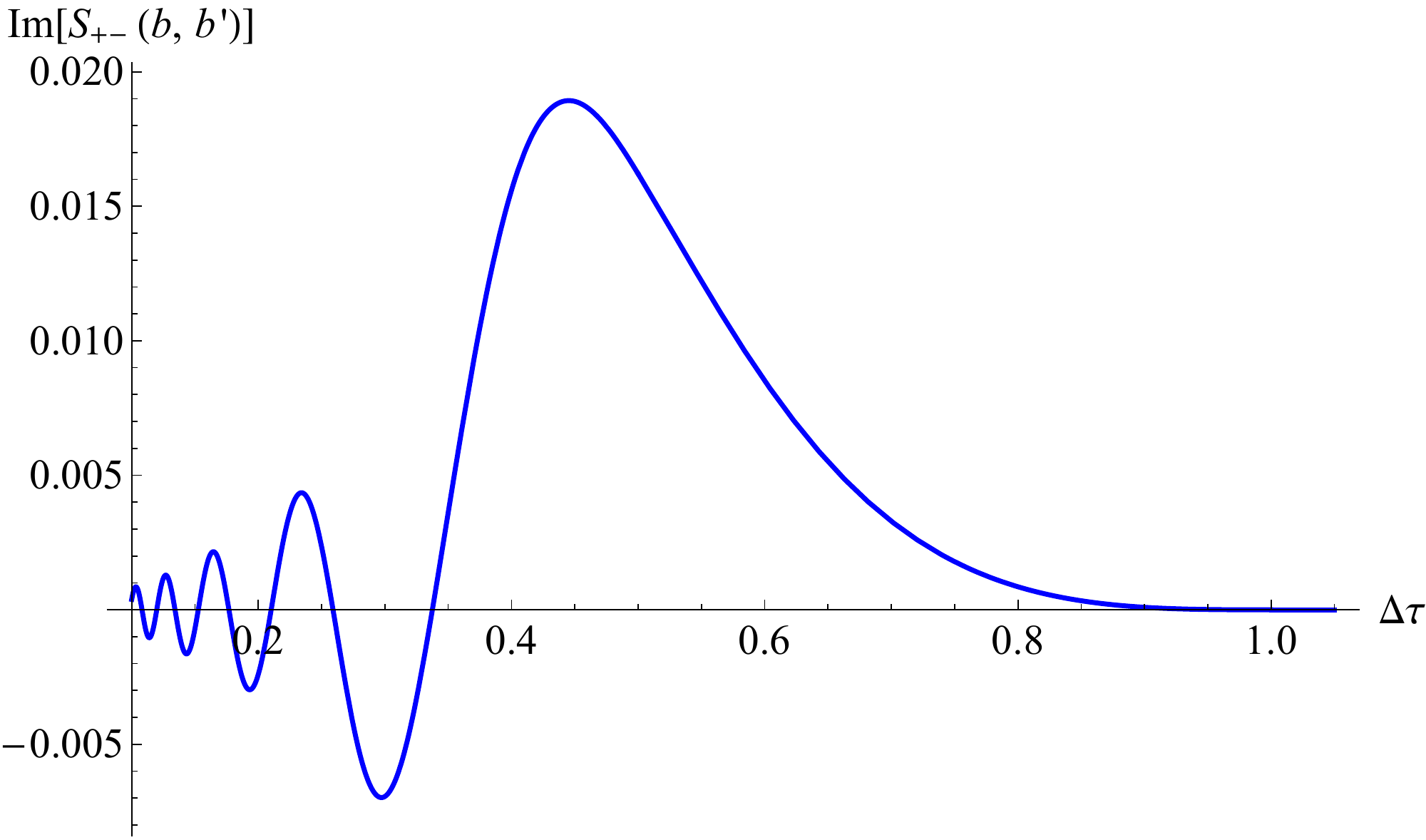}
        \caption{}
        \label{fig:orthoim}
    \end{subfigure}
    \caption{Plot of (a) real part, and (b) imaginary part of $S_{+-}(b,b')$ given by equation (\ref{qubitorthexact}) as a function of measurement time $\Delta \tau$ at parameter values $\lambda = 4$ $\alpha = 1$, $\kappa_0 = 0.25$, $g_0 = 2$, $b_0 = 1$, $b = 1.5$, and $b' = -2$. The vanishing of the real and imaginary parts of $S_{+-}(b,b')$ for $\Delta \tau \geq 1.00$ implies the exact orthogonality of the pointer states at those times. Evaluation of the $S_{+-}(b,b')$ by quadratures leads to the same plots as above.}\label{fig:qubitorthog}
\end{figure}

Then, the pointer states are $\rho_+$ and $\rho_{-}$, which are given by
\begin{equation}
\rho_{\pm} = \frac{C(\Delta \tau)}{b_0 \pi} \int_{-\infty}^{\infty} \mathrm{d}b \int_{-\infty}^{\infty} \mathrm{d}b'\, \frac{\sin[\frac{2\lambda \kappa_0}{\alpha g_0 \Delta \tau}(b-b')]}{(b-b')} \frac{\sin(b_0 b)}{b}\,  \frac{\sin(b_0 b')}{b'}\, \mathrm{e}^{\pm \frac{i \lambda}{2}(b-b')}\, |b\rangle \langle b'|,
\end{equation}
where $C(\Delta \tau) = \alpha g_0 \Delta \tau/2 \lambda \kappa_0$. In order to see how the pointer states $\rho_+$ and $\rho_-$ become exactly orthogonal, we investigate the function $S_{+-}(b,b')$ given by
\begin{equation}\label{spm}
S_{+-}(b,b') = \frac{C^2(\Delta \tau)}{b_0 \pi} \int_{-\infty}^{\infty} \mathrm{d}b'' \, \mathrm{e}^{-i\lambda b''} \, \frac{\sin^2(b_0 b'')}{b''^2} \frac{\sin[\frac{2 \lambda \kappa_0}{\alpha g_0 \Delta \tau}(b-b'')]}{(b-b'')} \, \frac{\sin[\frac{2 \lambda \kappa_0}{\alpha g_0 \Delta \tau}(b''-b')]}{(b''-b')},
\end{equation}
and obtain a condition for the vanishing of the $S_{+-}(b,b')$. It can be shown that the product of the three cardinal sine functions in equation (\ref{spm}) has a complex plane extension that is exponential of type $2b_0 + 4\lambda \kappa_0/\alpha g_0 \Delta \tau$. It follows from Lemma \ref{exptype} that $S_{+-}(b,b')$ vanishes provided the conditions $\lambda > 2b_0$ and  $\Delta \tau > \Delta \tau_O$ are satisfied, with the minimum $\Delta \tau_O = 4 \lambda \kappa_0 /\alpha g_0 (\lambda - 2b_0)$ the corresponding orthogonality time. However, we show from the closed form of $S_{+-}(b,b')$ that the pointer states become exactly orthogonal under the conditions $\Delta \tau \geq \Delta \tau_O$ and $\lambda > 2b_0$. With the same techniques employed in evaluating the integral for $I_{+-}(b,b')$, we can solve for the explicit form of $S_{+-}(b,b')$. The result is
\begin{equation}\label{qubitorthexact}
S_{+-}(b,b') = 
\begin{cases}
G_1(b,b') & \text{; $0 < \Delta \tau \leq \frac{4\lambda \kappa_0}{\alpha g_0 (\lambda + 2b_0)}$}, \\
\\
G_2(b,b') & \text{;$\frac{4\lambda \kappa_0}{\alpha g_0 (\lambda + 2b_0)} \leq \Delta \tau \leq \frac{4\kappa_0}{\alpha g_0}$},\\
\\
G_3(b,b') & \text{; $\frac{4\kappa_0}{\alpha g_0 } \leq \Delta \tau \leq \frac{4\lambda\kappa_0}{\alpha g_0 (\lambda -2b_0)}$},\\
\\
0 & \text{;$\Delta \tau \geq \frac{4\lambda\kappa_0}{\alpha g_0 (\lambda -2b_0)}$},
\end{cases}
\end{equation}
where
\begin{equation}\label{G1}
G_1(b,b') = \frac{i C^2(\Delta \tau)}{2b_0}  \, \left[ \frac{\mathrm{e}^{-i \lambda b' - \frac{2i \lambda \kappa_0}{\alpha g_0 \Delta \tau}(b-b')}}{(b-b')} \frac{\sin^2(b_0 b')}{b'^2}
- \frac{\mathrm{e}^{-i \lambda b + \frac{2i \lambda \kappa_0}{\alpha g_0 \Delta \tau}(b-b')}}{(b-b')} \frac{\sin^2(b_0 b)}{b^2} \right],
\end{equation}

\begin{align}\label{G2}
G_2(b,b') = \frac{C^2(\Delta \tau)}{8b_0} \left[\frac{f(b') \, \mathrm{e}^{-\frac{2i\lambda \kappa_0}{\alpha g_0 \Delta \tau}(b-b')}}{(b-b')} - \frac{f(b)\,\mathrm{e}^{\frac{2i\lambda \kappa_0}{\alpha g_0 \Delta \tau}(b-b')}}{(b-b')} -\frac{\phi_{+}(b,b')\,\mathrm{e}^{-\frac{2i\lambda \kappa_0}{\alpha g_0 \Delta \tau}(b+b')}}{b\,b'}\right],
\end{align}
and
\begin{align}\label{G3}
G_3(b,b') = \frac{C^2(\Delta \tau)}{8b_0} \left [\frac{h(b) \, \mathrm{e}^{\frac{2i\lambda \kappa_0}{\alpha g_0 \Delta \tau}(b-b')}}{(b-b')} 
- \frac{h(b')\,\mathrm{e}^{-\frac{2i\lambda \kappa_0}{\alpha g_0 \Delta \tau}(b-b')}}{(b-b')}
+\frac{\phi_{-}(b,b') \,\mathrm{e}^{-\frac{2i\lambda \kappa_0}{\alpha g_0 \Delta \tau}(b+b')}}{b\,b'} \right].
\end{align}
In equations (\ref{G1})-(\ref{G3}), we have the following shorthand notations: $f(x) = i\, \mathrm{e}^{-i\lambda x} [4 \sin^2(b_0x)\,+\,\mathrm{e}^{-2ib_0x}]/x^2$, $h(x) = i\,\mathrm{e}^{-i(\lambda - 2b_0)x}/x^2 $, and $\phi_{\pm}(b,b') = \pm 2b_0 + \lambda + i(b+b')/bb' - 4\lambda \kappa_0/\alpha g_0 \Delta \tau$. This closed form of $S_{+-}(b,b')$ holds under the condition $\lambda > 2b_0$. Thus the pointer states become exactly orthogonal for $\Delta \tau \geq 4\lambda \kappa_0/\alpha g_0 (\lambda - 2b_0)$ and provided that $\lambda > 2b_0$. Figure \ref{fig:qubitorthog} shows the plot of the real and imaginary parts of $S_{+-}(b,b')$ as functions of time $\Delta \tau$. Here we find that both $\mathrm{Re}[S_{+-}(b,b')]$ and $\mathrm{Im}[S_{+-}(b,b')]$  vanish for $\Delta \tau \geq 1.00$, with the minimum exactly equal to the orthogonality time $\Delta \tau_O = 1.00$  given the assumed values of the relevant parameters. By comparing the decoherence and orthogonality times, we have confirmed in this example that exact decoherence occurs at an earlier time than the time required for the pointer states to become exactly orthogonal, i.e. $\Delta \tau_O > \Delta \tau_D$. The possible outcomes of measuring $S_z$ are unambiguous at measurement times greater than or equal the orthogonality time. 

Now we show that when the orthogonality condition $\lambda> 2b_0$ holds and at $\Delta \tau = \Delta \tau_O$, the supports of the probability densities of the outcomes computed from the pointer states $\rho_+$ and $\rho_-$ do not overlap at orthogonality time. Since the output observable of the pointer is its momentum, let $|s\rangle$ be the pointer momentum eigenstate and $s$ be the corresponding momentum variable. We compute for the normalized probability densities $\langle s|\rho_{+} |s\rangle_{\Delta \tau = \Delta \tau_O}$ and $\langle s| \rho_-|s\rangle_{\Delta \tau = \Delta \tau_O}$ in the pointer momentum representation and evaluate these expressions at time $\Delta \tau$ equal to the orthogonality time. Here, there are three possible cases at which these terms have the following forms: when (i) $\lambda > 4b_0$, we have 
\begin{equation}
\langle s|\rho_{+} |s\rangle_{\Delta \tau = \Delta \tau_O} = 
\begin{cases}
0 & \text{; $s \leq 0,$}\\
\\
\frac{s}{2 b_0 \hbar^2 (\lambda - 2b_0)} & \text{; $0 \leq s \leq 2b_0\hbar,$}\\
\\
\frac{1}{(\lambda - 2b_0)\hbar} & \text{; $2b_0\hbar \leq s \leq (\lambda - 2b_0)\hbar,$}\\
\\
\frac{(\lambda - \frac{s}{\hbar})}{2 b_0 \hbar (\lambda - 2b_0)} & \text{; $(\lambda - 2b_0)\hbar \leq s \leq \lambda \hbar.$}\\
\\
0 & \text{; $s \geq \lambda \hbar,$}
\end{cases}
\end{equation}
and,
\begin{equation}
\langle s|\rho_{-} |s\rangle_{\Delta \tau = \Delta \tau_O} = 
\begin{cases}
0 & \text{; $s \geq 0,$}\\
\\
-\frac{s}{2 b_0 \hbar^2 (\lambda - 2b_0)} & \text{; $-2b_0\hbar \leq s \leq 0,$}\\
\\
\frac{1}{(\lambda - 2b_0)\hbar} & \text{; $-(\lambda - 2b_0)\hbar \leq s \leq -2b_0\hbar,$}\\
\\
\frac{(\lambda + \frac{s}{\hbar})}{2 b_0 \hbar (\lambda - 2b_0)} & \text{; $- \lambda \hbar \leq s \leq -(\lambda - 2b_0)\hbar,$}\\
\\
0 & \text{; $s \leq -\lambda \hbar.$}
\end{cases}
\end{equation}
\\
\noindent When (ii) $2b_0 <\lambda < 4b_0$,
\begin{equation}
\langle s|\rho_{+} |s\rangle_{\Delta \tau = \Delta \tau_O} = 
\begin{cases}
0 & \text{; $s \leq 0,$}\\
\\
\frac{s}{2 b_0 \hbar^2 (\lambda - 2b_0)} & \text{; $0 \leq s \leq (\lambda - 2b_0)\hbar,$}\\
\\
\frac{1}{2b_0\hbar} & \text{; $(\lambda - 2b_0)\hbar \leq s \leq 2 b_0 \hbar,$}\\
\\
\frac{(\lambda - \frac{s}{\hbar})}{2 b_0 \hbar (\lambda - 2b_0)} & \text{; $2 b_0 \hbar  \leq s \leq \lambda \hbar,$}\\
\\
0 & \text{; $s \geq \lambda \hbar,$}
\end{cases}
\end{equation}
and
\begin{equation}
\langle s|\rho_{-} |s\rangle_{\Delta \tau = \Delta \tau_O} = 
\begin{cases}
0 & \text{; $s \geq 0,$}\\
\\
-\frac{s}{2 b_0 \hbar^2 (\lambda - 2b_0)} & \text{; $-(\lambda - 2b_0)\hbar \leq s \leq 0,$}\\
\\
\frac{1}{2b_0\hbar} & \text{; $-2 b_0 \hbar \leq s \leq -(\lambda - 2b_0)\hbar,$}\\
\\
\frac{(\lambda + \frac{s}{\hbar})}{2 b_0 \hbar (\lambda - 2b_0)} & \text{; $- \lambda \hbar \leq s \leq -2 b_0 \hbar,$}\\
\\
0 & \text{; $s \leq -\lambda \hbar,$}
\end{cases}
\end{equation}
\\
Lastly, when (iii) $\lambda = 4b_0$,
\begin{equation}
\langle s|\rho_{+} |s\rangle_{\Delta \tau = \Delta \tau_O} = 
\begin{cases}
0 & \text{; $s \leq 0,$}\\
\\
\frac{4s}{\lambda^2 \hbar^2} & \text{; $0 \leq s \leq \frac{\lambda \hbar}{2},$}\\
\\
\frac{4 (\lambda - \frac{s}{\hbar})}{\lambda^2 \hbar} & \text{; $\frac{\lambda \hbar}{2} \leq s \leq \lambda \hbar,$}\\
\\
0 & \text{; $s \geq \lambda \hbar,$}
\end{cases}
\end{equation}
and
\begin{equation}
\langle s|\rho_{-} |s\rangle_{\Delta \tau = \Delta \tau_O} = 
\begin{cases}
0 & \text{; $s \geq 0,$}\\
\\
-\frac{4s}{\lambda^2 \hbar^2} & \text{; $-\frac{\lambda \hbar}{2} \leq s \leq 0,$}\\
\\
\frac{4(\lambda + \frac{s}{\hbar})}{\lambda^2 \hbar} & \text{; $- \lambda \hbar \leq s \leq -\frac{\lambda \hbar}{2} ,$}\\
\\
0 & \text{; $s \leq -\lambda \hbar,$}
\end{cases}
\end{equation}
\begin{figure*}[t]
    \centering
    \begin{subfigure}[b]{0.31\textwidth}
        \includegraphics[width=\textwidth]{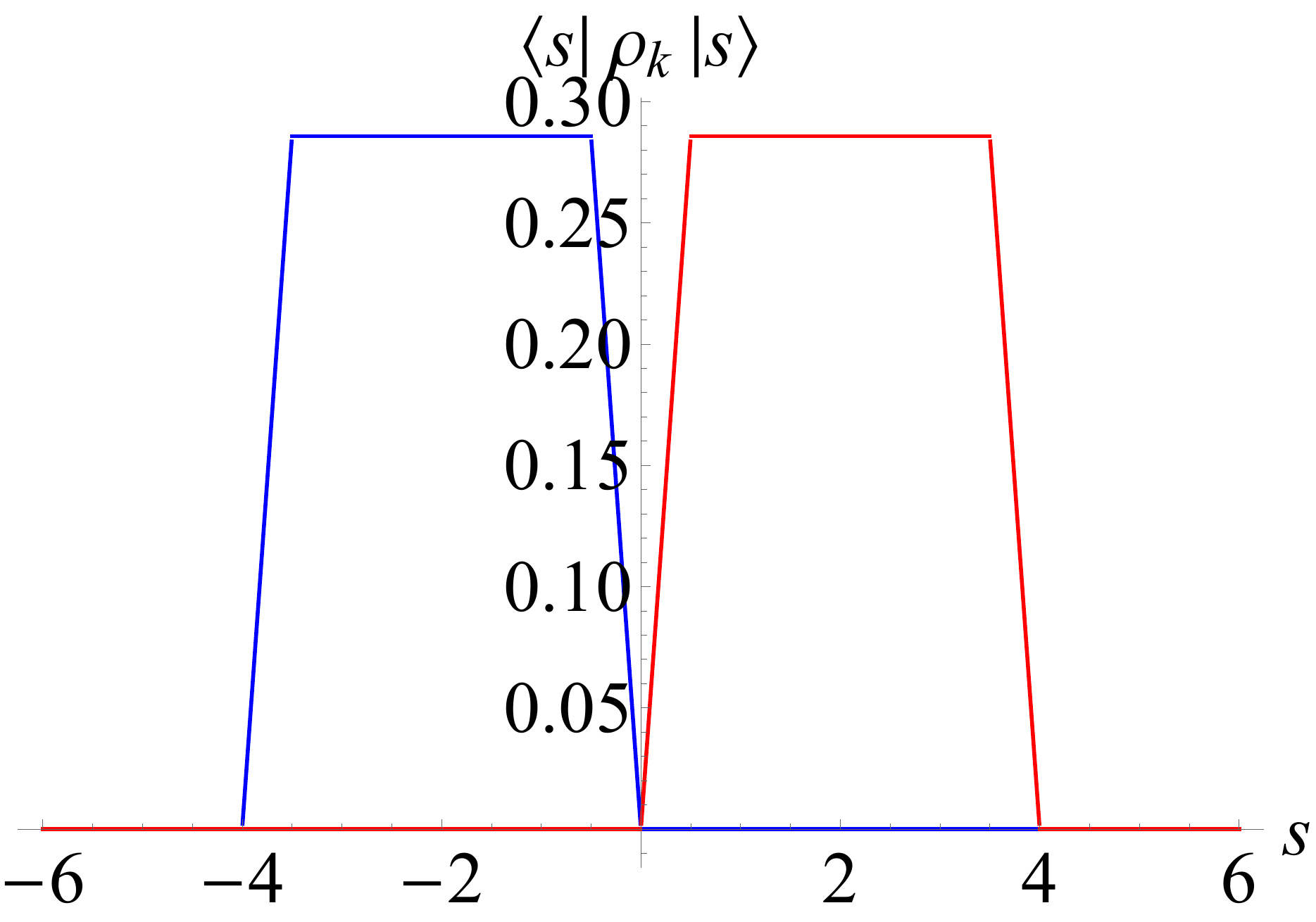}
        \caption{}
        \label{fig:probdenqubitc1}
    \end{subfigure}
    ~ 
    \begin{subfigure}[b]{0.31\textwidth}
        \includegraphics[width=\textwidth]{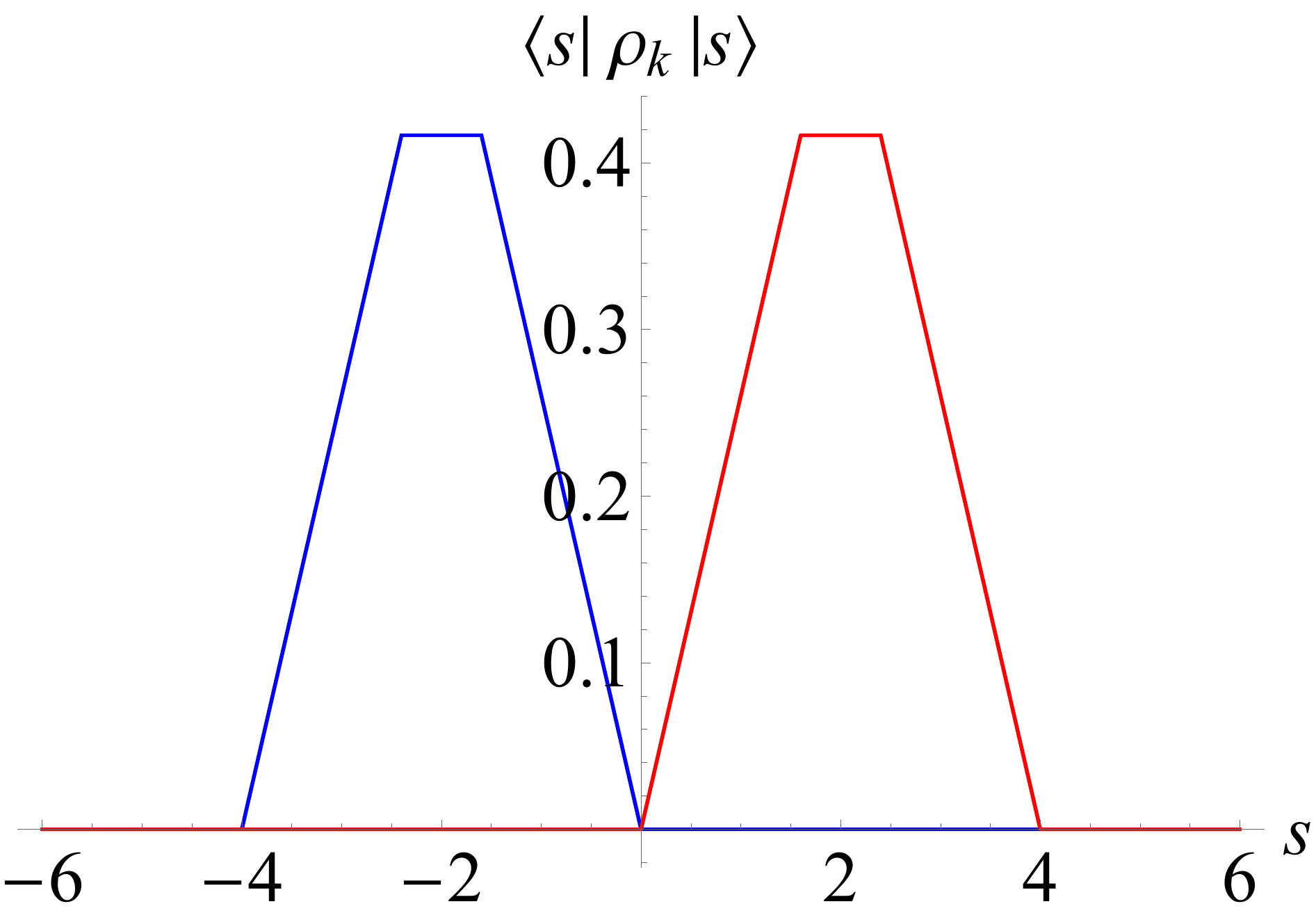}
        \caption{}
        \label{fig:probdenqubitc2}
    \end{subfigure}
    ~ 
    \begin{subfigure}[b]{0.31\textwidth}
        \includegraphics[width=\textwidth]{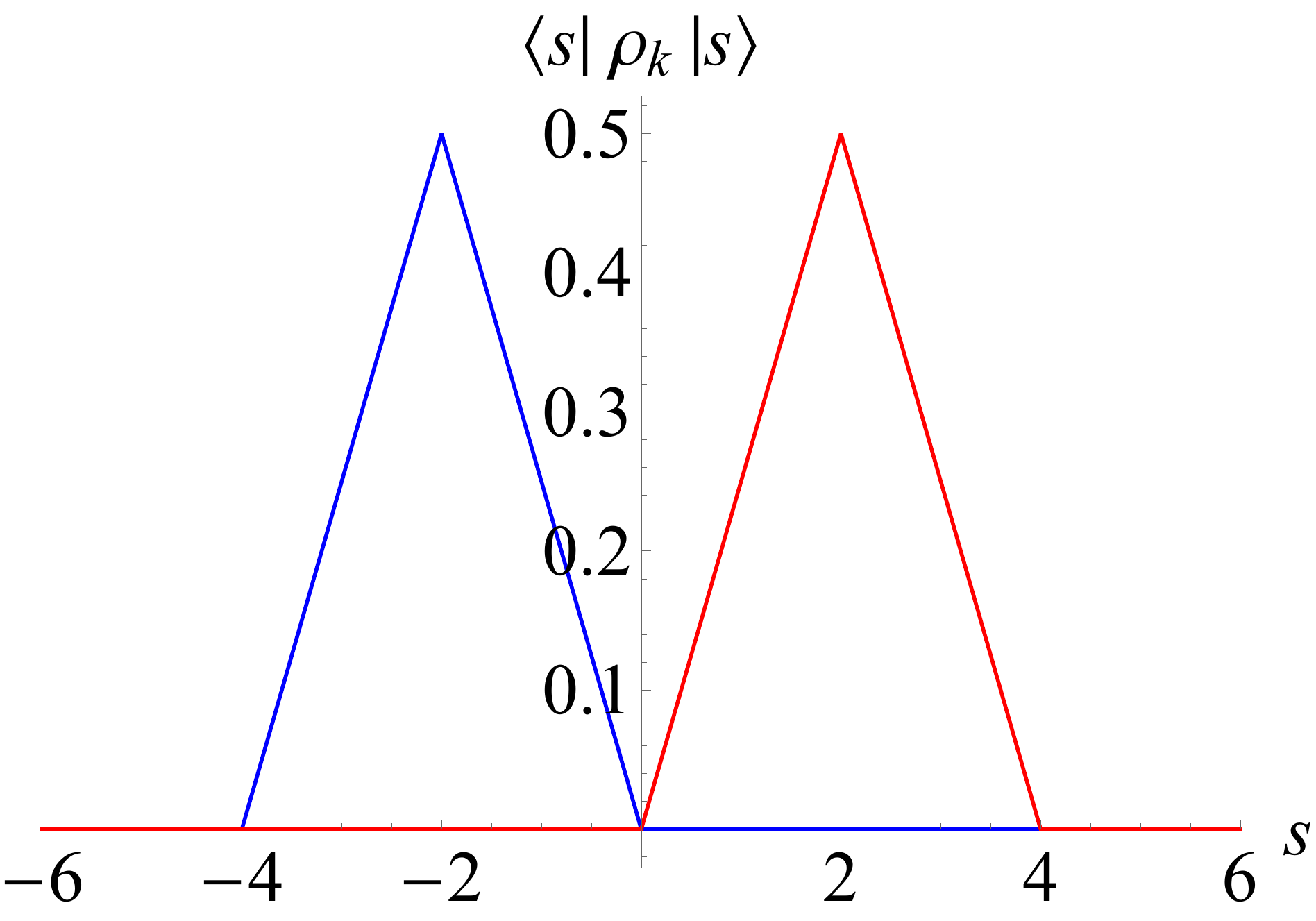}
        \caption{}
        \label{fig:probdenqubitc3}
    \end{subfigure}
    \caption{Plot of the diagonal terms $\langle s|\rho_{+} |s\rangle_{\Delta \tau = \Delta \tau_O}$ (shown by the red plot) and $\langle s|\rho_{-} |s\rangle_{\Delta \tau = \Delta \tau_O}$ (shown by the blue plot) versus the pointer momentum $s$ for the following cases: (a) $\lambda > 4b_0$, (b) $\lambda < 4b_0$, and (c) $\lambda = 4b_0.$ In all cases, we have let $\lambda = 4$ and $\hbar = 1$ while we let $b_0 = 0.25$ in case (a), $b_0 = 1.2$ in case (b), and $b_0 = 1$ in case (c).}\label{fig:qubitprobden}
\end{figure*}
\noindent In each case, $\langle s|\rho_{+} |s\rangle_{\Delta \tau = \Delta \tau_O}$ has a support in the interval $[0, \lambda \hbar]$. Moreover, it can be shown in each case that $\langle s|\rho_{+} |s\rangle_{\Delta \tau = \Delta \tau_O}$ is symmetric and maximum at $s = \lambda \hbar/2$. On the other hand, $\langle s|\rho_{-}|s\rangle_{\Delta \tau = \Delta \tau_O}$ has a support in the interval $[-\lambda \hbar, 0]$ and it is symmetric and maximum at $s = -\lambda \hbar/2$. Clearly, the supports of the $\langle s|\rho_{+} |s\rangle_{\Delta \tau = \Delta \tau_O}$ and $\langle s|\rho_{-}|s\rangle_{\Delta \tau = \Delta \tau_O}$ do not overlap. These results imply that a projective measurement of the observable $S_z$ of the spin-1/2 system can be realized through the apparatus pointer at $\Delta \tau = \Delta \tau_O$. We show the plots of $\langle s|\rho_{+} |s\rangle_{\Delta \tau = \Delta \tau_O}$ and $\langle s|\rho_{-}|s\rangle_{\Delta \tau = \Delta \tau_O}$ as functions of $s$ for each of the mentioned cases  and for specific parameter values in Figure \ref{fig:qubitprobden} (as shown in red and blue plots, respectively).

\subsection{Quantum Harmonic Oscillator}
\begin{figure}[t]
    \centering
    \captionsetup{width=.8\linewidth}
    \begin{subfigure}[b]{0.45\textwidth}
        \includegraphics[width=\textwidth]{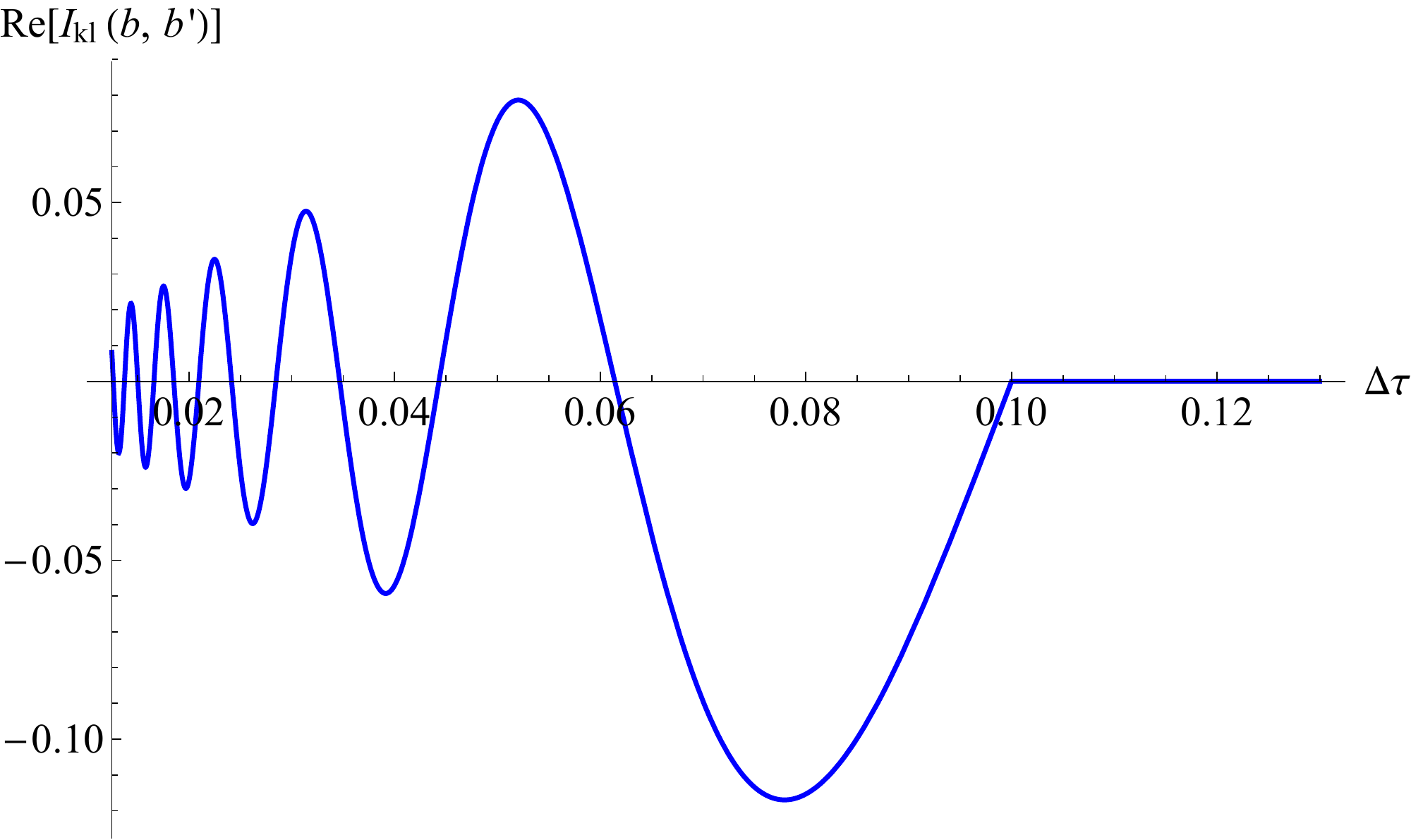}
        \caption{}
        \label{fig:CohRealHO}
    \end{subfigure}
    ~ 
    \begin{subfigure}[b]{0.45\textwidth}
        \includegraphics[width=\textwidth]{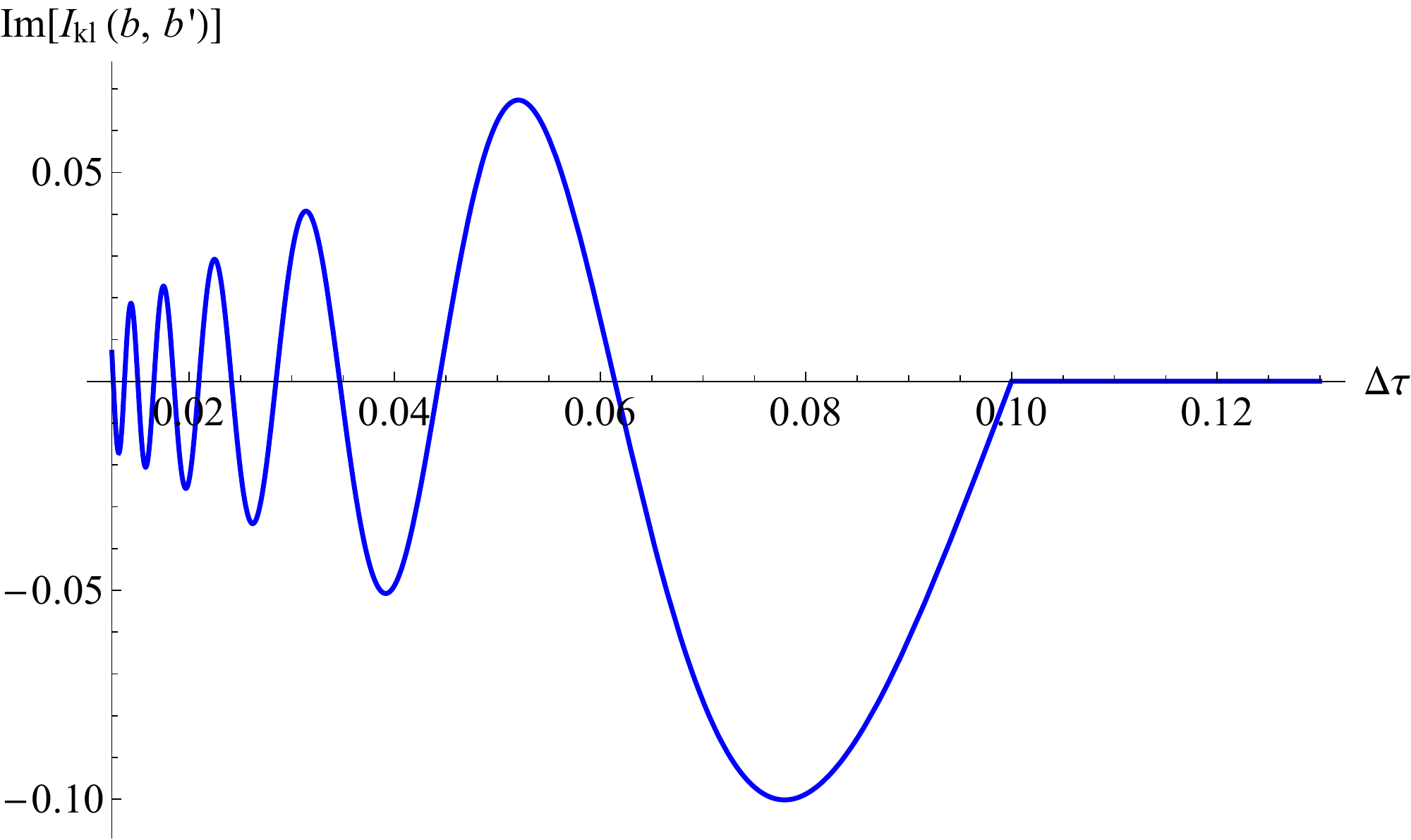}
        \caption{}
        \label{fig:cohImHO}
    \end{subfigure}
    \caption{Plot of (a) real part, and (b) imaginary part of the decoherence factors $I_{k \neq l}(b,b')$ given by equation (\ref{HOcohexact}) as a function of interaction time $\Delta \tau$ for the measurement of energy observable of a quantum harmonic oscillator, with parameter values $\lambda = 2$, $\omega = 2.5$, $\alpha = 2$, $\kappa_0 = 0.5$, $g_0 = 2$, $b = 1$, $b' = 2$, $k=2$ and $l=1$. Both real and imaginary parts of $I_{k \neq l}(b,b')$ vanish at and beyond $\Delta \tau = 0.10$. Evaluation of the $I_{k \neq l}(b,b')$ by quadratures leads to the same plots as above.}\label{fig:HOdeco}
\end{figure}

We consider the measurement of the energy observable of the quantum harmonic oscillator with mass $m$ and angular frequency $\omega$ by using the measurement model of \cite{galapon1}. Despite the fact that the quantum harmonic oscillator has an underlying infinite-dimensional Hilbert space, the oscillator's energy observable represented by its Hamiltonian is discrete, non-degenerate, and has eigenvalues and corresponding eigenstates given by the eigenvalue equation $H |\varphi_k\rangle = (k + 1/2)\hbar \omega|\varphi_k\rangle$, where $k = 0,1,2,...$ is the oscillator's quantum number. Also, the minimum difference between the possible energy values of the harmonic oscillator is equal to $\hbar \omega$ \cite{griffiths}. Likewise, we assume momentum-limited initial states of the probe and pointer which are given by equations (\ref{ipr}) and (\ref{ipo}). Then the corresponding closed-form expression for decoherence factors $I_{k \neq l}(b,b')$ is 

\begin{equation}\label{HOcohexact}
I_{k \neq l}(b,b') = \frac{i\alpha g_0 \Delta \tau}{4 \lambda \kappa_0}\, V_{kl}(b,b')\,\Theta(2\kappa_0 - \omega \alpha g_0 |k-l| \Delta \tau), 
\end{equation}
where
\begin{equation}
V_{kl}(b,b') = \frac{\,\mathrm{e}^{-2i\lambda \left[\frac{\kappa_0}{\alpha g_0 \Delta \tau}(b-b') + \omega |k-l| b' \right]}}{b-b'} - \frac{\mathrm{e}^{2i\lambda \left[\frac{\kappa_0}{\alpha g_0 \Delta \tau}(b-b') - \omega |k-l| b \right]}}{b-b'}.
\end{equation}
Note that $k, l = 0, 1, 2, ...$. It is easy to see that all $I_{k \neq l}(b,b')$'s vanish for $\Delta \tau \geq \Delta \tau_D'$, where $\Delta \tau_D' = 2\kappa_0/\alpha g_0 \omega$ is the corresponding decoherence time. Given this condition, exact decoherence can be achieved at an earlier time for large $\omega$.  Figure \ref{fig:HOdeco} shows the plot of the real and imaginary parts of $I_{k \neq l}(b,b')$ as functions of time $\Delta \tau$. Both real and imaginary parts oscillate with increasing amplitude in time but then vanish for $\Delta \tau \geq 0.10$, where the minimum value is the decoherence time $\Delta \tau_D' = 0.10$. 

Moreover, in considering the orthogonality of the pointer states $\rho_k$'s, we evaluate the function $S_{k \neq l}(b,b')$. This is found to have the explicit form
\begin{equation}\label{HOorthexact}
S_{k \neq l}(b,b') = 
\begin{cases}
K_1(b,b') & \text{; $0 < \Delta \tau \leq \frac{4\lambda \kappa_0}{\alpha g_0 [\lambda \omega |k-l| + 2b_0]}$}, \\
\\
K_2(b,b') & \text{;$\frac{4\lambda \kappa_0}{\alpha g_0 [\lambda \omega (k-l) + 2b_0]} \leq \Delta \tau \leq \frac{4\kappa_0}{\alpha g_0 \omega |k-l|}$},\\
\\
K_3(b,b') & \text{; $\frac{4\kappa_0}{\alpha g_0 \omega (k-l)} \leq \Delta \tau \leq \frac{4\lambda\kappa_0}{\alpha g_0 [\lambda \omega |k-l| -2b_0]}$},\\
\\
0 & \text{;$\Delta \tau \geq \frac{4\lambda\kappa_0}{\alpha g_0 [\lambda \omega |k-l|-2b_0]}$},
\end{cases}
\end{equation}

\noindent where 
\begin{equation}\label{K1}
K_1(b,b') = \frac{i C^2(\Delta \tau)}{2b_0}  \, \left[ \frac{\mathrm{e}^{-i \lambda \omega |k-l| b' - \frac{2i \lambda \kappa_0}{\alpha g_0 \Delta \tau}(b-b')}}{(b-b')} \frac{\sin^2(b_0 b')}{b'^2}
- \frac{\mathrm{e}^{-i \lambda \omega |k-l| b + \frac{2i \lambda \kappa_0}{\alpha g_0 \Delta \tau}(b-b')}}{(b-b')} \frac{\sin^2(b_0 b)}{b^2} \right],
\end{equation}

\begin{equation}\label{K2}
K_2(b,b') = \frac{C^2(\Delta \tau)}{8b_0} \left[\frac{\tilde{f}(b') \, \mathrm{e}^{-\frac{2i\lambda \kappa_0}{\alpha g_0 \Delta \tau}(b-b')}}{(b-b')} 
- \frac{\tilde{f}(b)\,\mathrm{e}^{\frac{2i\lambda \kappa_0}{\alpha g_0 \Delta \tau}(b-b')}}{(b-b')} -\frac{\tilde{\phi}_{+}(b,b')\,\mathrm{e}^{-\frac{2i\lambda \kappa_0}{\alpha g_0 \Delta \tau}(b+b')}}{b\,b'}\right],
\end{equation}
and
\begin{equation}\label{K3}
K_3(b,b') = \frac{C^2(\Delta \tau)}{8b_0} \left [\frac{\tilde{h}(b) \, \mathrm{e}^{\frac{2i\lambda \kappa_0}{\alpha g_0 \Delta \tau}(b-b')}}{(b-b')} 
- \frac{\tilde{h}(b')\,\mathrm{e}^{-\frac{2i\lambda \kappa_0}{\alpha g_0 \Delta \tau}(b-b')}}{(b-b')}
+\frac{\tilde{\phi}_{-}(b,b') \,\mathrm{e}^{-\frac{2i\lambda \kappa_0}{\alpha g_0 \Delta \tau}(b+b')}}{b\,b'} \right].
\end{equation}

\begin{figure}[t]
    \centering
    \captionsetup{width=.8\linewidth}
    \begin{subfigure}[b]{0.48\textwidth}
        \includegraphics[width=\textwidth]{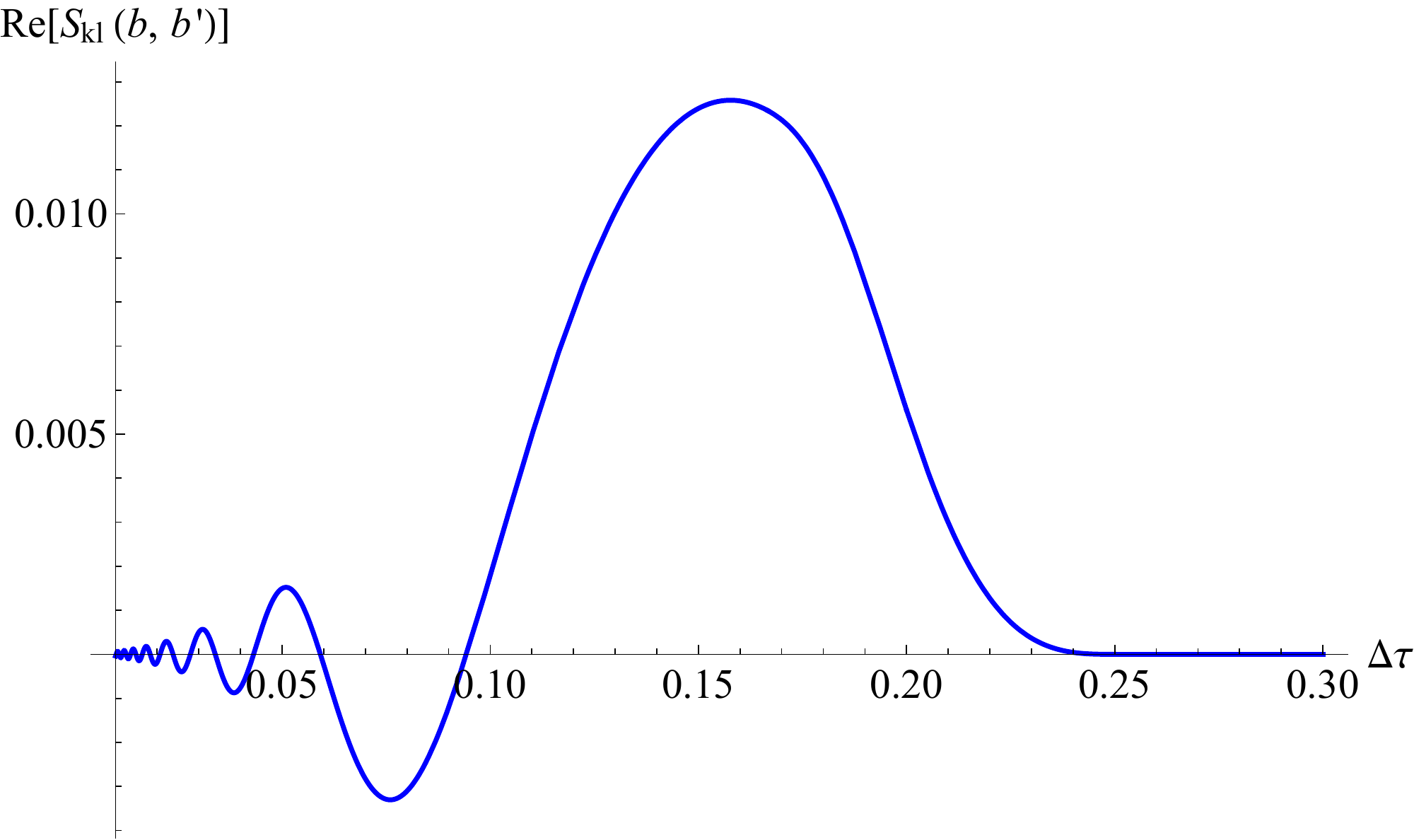}
        \caption{}
        \label{fig:OrthogHOReal}
    \end{subfigure}
    ~ 
    \begin{subfigure}[b]{0.48\textwidth}
        \includegraphics[width=\textwidth]{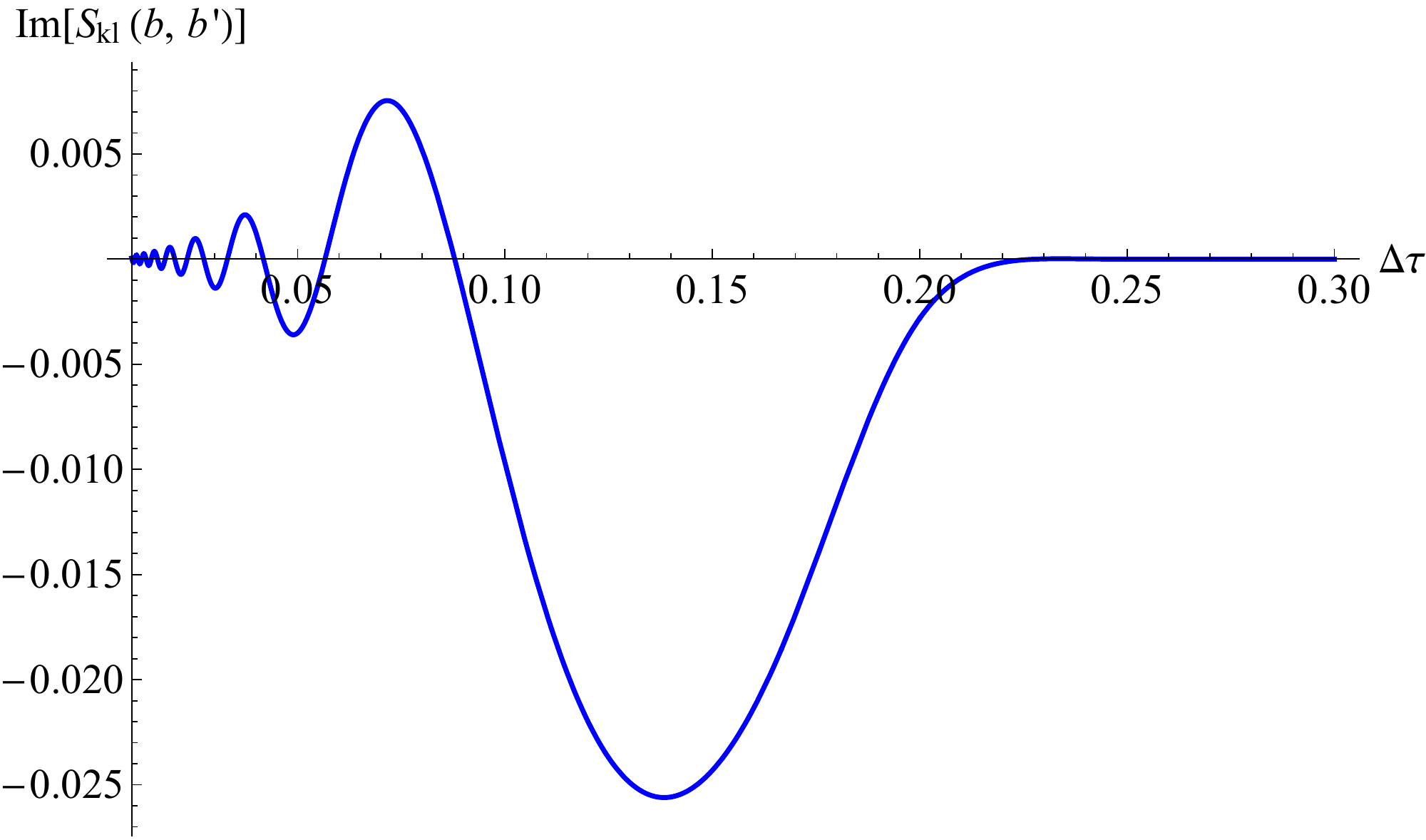}
        \caption{}
        \label{fig:OrthogHOIm}
    \end{subfigure}
    \caption{Plot of (a) real part, and (b) imaginary part of $S_{k \neq l}(b,b')$ given by equation (\ref{HOorthexact}) as a function of interaction time $\Delta \tau$ at parameter values $\lambda = 2$, $\omega = 2.5$, $\alpha = 2$, $\kappa_0 = 0.5$, $g_0 = 2$, $b_0 = 0.5$, $b = 1$, $b' = 2$, $k=2$ and $l=1$. The vanishing of $S_{k \neq l}(b,b')$ at $\Delta \tau \geq 0.25$ implies the exact orthogonality of the pointer states at those times. Evaluation of the $S_{k \neq l}(b,b')$ by quadratures leads to the same plots as above. }\label{fig:HOorthog}
\end{figure}

\noindent In equations (\ref{K1})-(\ref{K3}), we have the following shorthand notations: $\tilde{f}(x) = i\, \mathrm{e}^{-i\lambda \omega |k-l| x} [4 \sin^2(b_0x)\,+\,\mathrm{e}^{-2ib_0x}]/x^2$, $\tilde{h}(x) = i\,\mathrm{e}^{-i[\lambda \omega |k-l| - 2b_0]x}/x^2 $, and $\tilde{\phi}_{\pm}(b,b') =  \pm 2b_0 \,+\, \lambda \omega |k-l| + i(b+b')/bb' \,-\, 4\lambda \kappa_0/\alpha g_0 \Delta \tau$. This closed form of $S_{k \neq l}(b,b')$ holds provided that $\lambda \omega |k-l| - 2b_0 >0$. Thus the pointer states become exactly orthogonal for $\Delta \tau \geq  \Delta \tau_O'$, where $\Delta \tau_O'$ is the corresponding orthogonality time $ \Delta \tau_O' = 4\lambda \kappa_0/\alpha g_0 (\lambda \omega - 2b_0)$, and $\lambda \omega > 2b_0$. Figure \ref{fig:HOorthog} shows the plot of the real and imaginary parts of $S_{k \neq l}(b,b')$ as functions of time $\Delta \tau$. Both real and imaginary parts vanish for $\Delta \tau \geq \Delta \tau_O' = 0.25$. By comparing the decoherence and orthogonality times in this example, we have confirmed that $\Delta \tau_O' > \Delta \tau_D'$. 

\begin{figure}[t!]
    \centering
    \begin{subfigure}[b]{0.335\textwidth}
        \includegraphics[width=\textwidth]{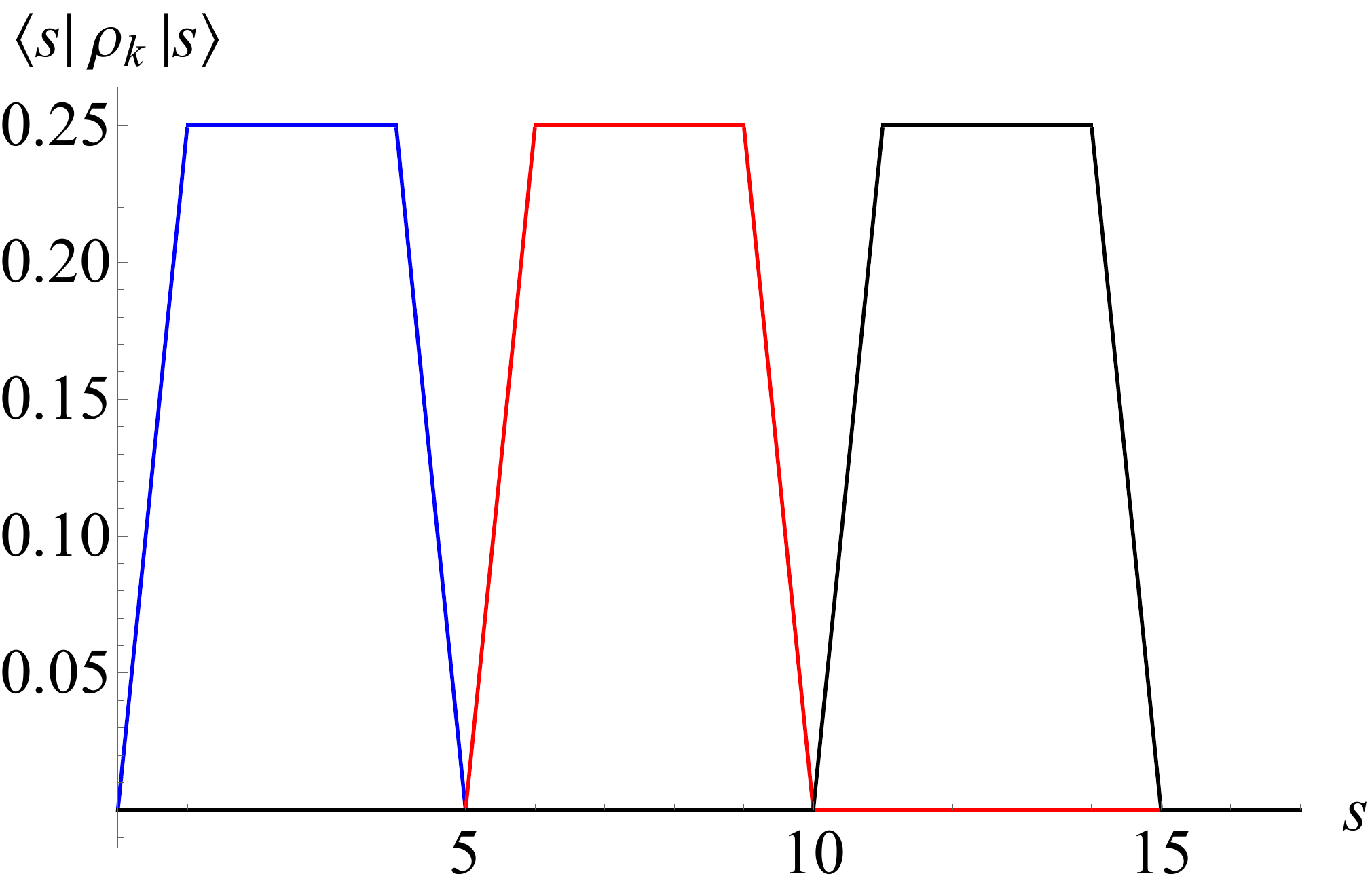}
        \caption{}
        \label{fig:probdenHOc1}
    \end{subfigure}
    ~ 
    \begin{subfigure}[b]{0.31\textwidth}
        \includegraphics[width=\textwidth]{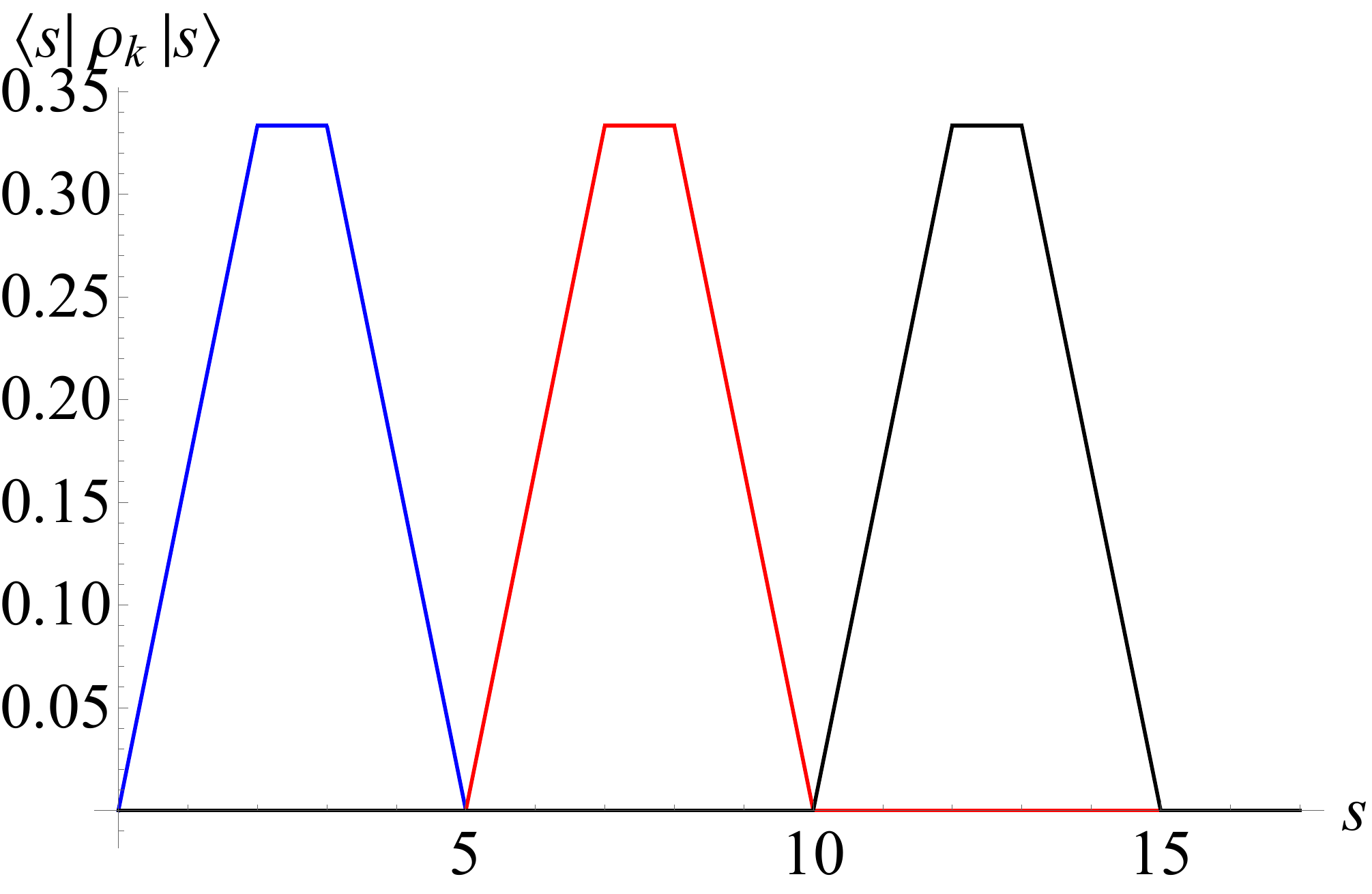}
        \caption{}
        \label{fig:probdenHOc2}
    \end{subfigure}
    ~ 
    \begin{subfigure}[b]{0.31\textwidth}
        \includegraphics[width=\textwidth]{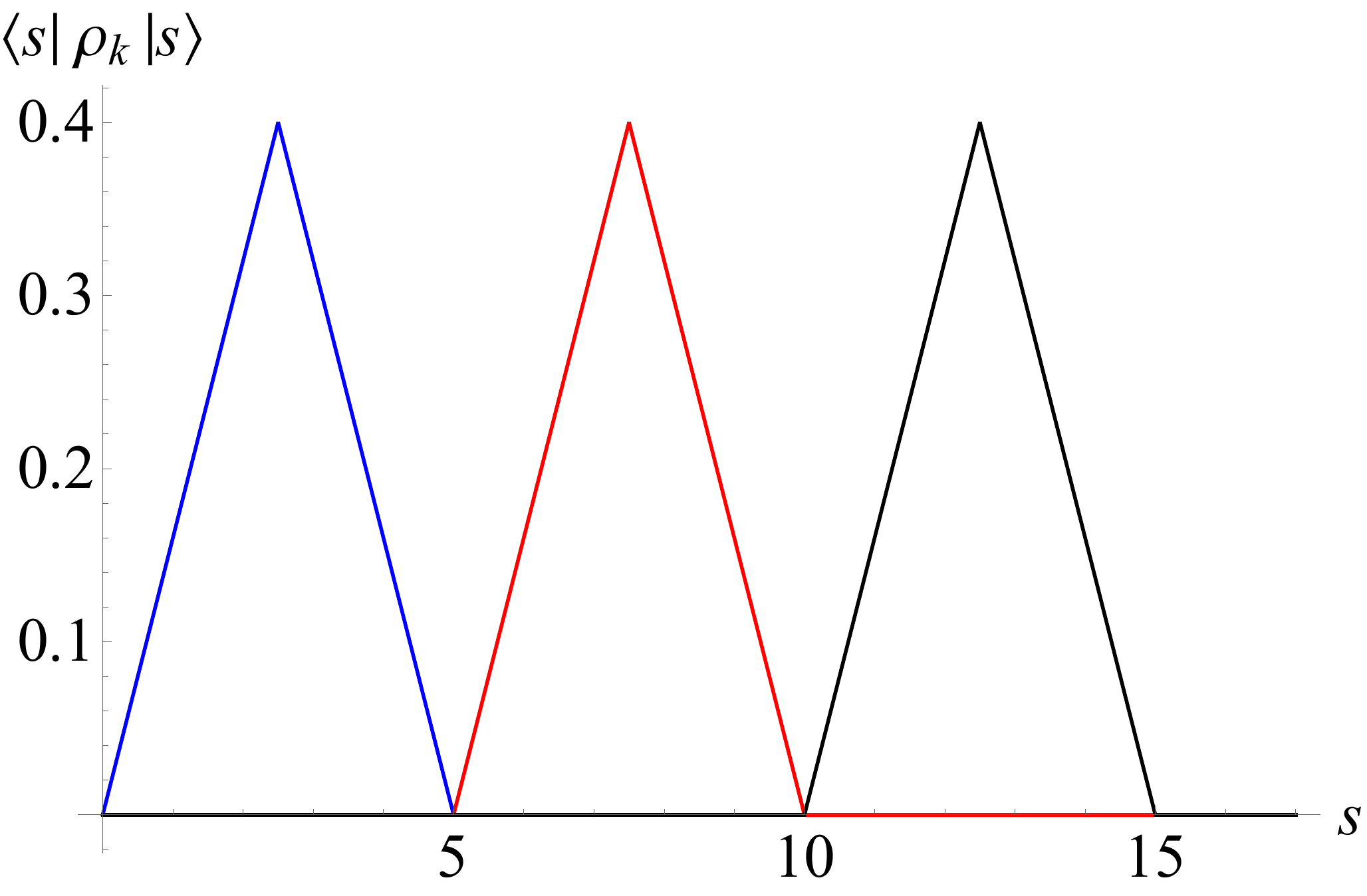}
        \caption{}
        \label{fig:probdenHOc3}
    \end{subfigure}
    \caption{Plot of $\langle s|\rho_k|s\rangle_{\Delta \tau = \Delta \tau_O'}$ as a function of the pointer momentum $s$ for $k = 0$ (blue), $k=1$ (red), and $k=2$ (black) for the  following cases: (a) $\lambda \omega > 4b_0$, (b) $\lambda \omega < 4b_0$, and (c) $\lambda \omega =4b_0.$ Here we set $\lambda = 2$, $\omega = 2.5$ and $\hbar =1$ in all cases while we let $b_0 = 0.5$ in case (a), $b_0 = 1.5$ in case (b), and $b_0 = 1.25$ in case (c).}\label{fig:HOprobden}
\end{figure}

Similarly, we show that the supports of the probability densities computed from pointer states $\rho_k$'s are are non-overlapping at $\Delta \tau = \Delta \tau_O'$ and $\lambda \omega > 2b_0$ for different $k$. In the pointer momentum $s$-representation, it is found that the probability densities $\langle s|\rho_k|s\rangle_{\Delta \tau = \Delta \tau_O'}$'s take forms in the following three cases: (i) when $\lambda \omega > 4b_0$,
\begin{equation}
\langle s|\rho_k |s\rangle_{\Delta \tau = \Delta \tau_O'} = 
\begin{cases}
0 & \text{; $s \leq \lambda \omega k \hbar,$}\\
\\
\frac{(\frac{s}{\hbar} - \lambda \omega k)}{2 b_0 \hbar (\lambda \omega - 2b_0)} & \text{; $\lambda \omega k \hbar \leq s \leq (\lambda \omega k + 2b_0) \hbar,$}\\
\\
\frac{1}{(\lambda \omega - 2b_0)\hbar} & \text{; $(\lambda \omega k + 2b_0)\hbar \leq s \leq [\lambda \omega(k+1) - 2b_0]\hbar,$}\\
\\
\frac{(\lambda \omega (k+1) - \frac{s}{\hbar})}{2 b_0 \hbar (\lambda \omega - 2b_0)} & \text{; $[\lambda \omega(k+1) - 2b_0]\hbar \leq s \leq \lambda \omega(k+1) \hbar,$}\\
\\
0 & \text{; $s \geq \lambda \omega(k+1) \hbar,$}
\end{cases}
\end{equation}
\\
(ii) when $2b_0 < \lambda \omega < 4b_0$,
\begin{equation}
\langle s|\rho_k |s\rangle_{\Delta \tau = \Delta \tau_O'} = 
\begin{cases}
0 & \text{; $s \leq \lambda \omega k \hbar,$}\\
\\
\frac{(\frac{s}{\hbar} - \lambda \omega k)}{2 b_0 \hbar (\lambda \omega - 2b_0)} & \text{; $\lambda \omega k \hbar \leq s \leq [\lambda \omega(k+1) - 2b_0]\hbar,$}\\
\\
\frac{1}{2b_0\hbar} & \text{; $[\lambda \omega(k+1) - 2b_0]\hbar \leq s \leq (\lambda \omega k + 2b_0)\hbar,$}\\
\\
\frac{(\lambda \omega (k+1) - \frac{s}{\hbar})}{2 b_0 \hbar (\lambda \omega - 2b_0)} & \text{; $(\lambda \omega k + 2b_0)\hbar \leq s \leq \lambda \omega(k+1) \hbar,$}\\
\\
0 & \text{; $s \geq \lambda \omega(k+1) \hbar,$}
\end{cases}
\end{equation}
\\
and (iii) when $\lambda \omega = 4b_0$,
\begin{equation}
\langle s|\rho_k |s\rangle_{\Delta \tau = \Delta \tau_O'} = 
\begin{cases}
0 & \text{; $s \leq \lambda \omega k \hbar,$}\\
\\
\frac{4(\frac{s}{\hbar} - \lambda \omega k)}{\lambda^2 \omega^2 \hbar} & \text{; $\lambda \omega k \hbar \leq s \leq \lambda\hbar\omega(k+\frac{1}{2}),$}\\
\\

\frac{4(\lambda \omega (k+1) - \frac{s}{\hbar})}{\lambda^2 \omega^2 \hbar} & \text{; $\lambda\hbar\omega(k+\frac{1}{2}) \leq s \leq \lambda \omega(k+1) \hbar,$}\\
\\
0 & \text{; $s \geq \lambda \omega(k+1) \hbar.$}
\end{cases}
\end{equation}
In each case, $\langle s|\rho_k |s\rangle_{\Delta \tau = \Delta \tau_O'}$ has a compact support in the interval $[k\lambda \hbar \omega, (k+1)\lambda \hbar \omega]$ and is symmetric at $s = \lambda (k+1/2)\hbar \omega$. For different values of $k$, the support of the probability densities $\langle s|\rho_k |s\rangle_{\Delta \tau = \Delta \tau_O'}$ do not overlap. This implies that a projective measurement of the energy observable of the quantum harmonic oscillator can be realized through the apparatus pointer at $\Delta \tau = \Delta \tau_O'$. Figure \ref{fig:HOprobden} shows the plot of $\langle s|\rho_k |s\rangle_{\Delta \tau = \Delta \tau_O'}$ as functions of the pointer momentum $s$ for $k=0,1,2$ in each of the cases (i)-(iii). 

\section{\label{sec:level5}Summary and Conclusions}
In this paper, we reexamine the system-probe-pointer measurement model of \cite{galapon1} and confirm the joint state of the system-probe-pointer composite system by solving the von Neumann equation for the joint density matrix of the system, probe and pointer as a function of measurement time subject to the condition that the system, probe and pointer are initially prepared in an uncorrelated state and under the assumption that quantum dynamics is dictated by the measurement Hamiltonian given by equation (\ref{Hamiltonian}). Here, we are able to reproduce the same joint density matrix of the system-probe-pointer composite system obtained in \cite{galapon1} via the factorization method of the time evolution operator, hence we are able to verify the conditions for exact decoherence and exactly orthogonal pointer states by a rigorous treatment of the system-probe-pointer measurement model. We also demonstrate the exact decoherence and measurement process of the system-probe-pointer measurement model in measuring the $z$-component of spin observable of a spin-1/2 particle and energy observable of a quantum harmonic oscillator given specific momentum-limited initial states of the probe and pointer. In each example, we have calculated the closed forms of the decoherence factors $I_{k \neq l}(b, b')$ find that exact decoherence occurs for $\Delta \tau \geq \Delta \tau_D$, where $\Delta \tau_D$ is the corresponding decoherence time. Also we are able to compute in each example, the closed form expressions for the functions $S_{k\neq l}(b,b')$ that measures of orthogonality of pointer states, and we find that the pointer states become exactly orthogonal given the conditions $\lambda > 2b_0\hbar/a_0$ and $\Delta \tau \geq \Delta \tau_O$, where $\Delta \tau_O$ is the corresponding orthogonality time. We compare the decoherence and orthogonality times in each example and we have confirmed the relation $\Delta \tau_O > \Delta \tau_D$, which implies that exact decoherence occurs at an earlier time and followed by the exact orthogonality of pointer states. In each example, we also solve for the probability densities $\langle s|\rho_k|s\rangle$ in the pointer momentum space evaluated at orthogonality time and under the respective orthogonality condition $\lambda > 2b_0 \hbar /a_0$. In each example, we find that the support of $\langle s|\rho_k|s\rangle_{\Delta \tau = \Delta \tau_O}$ are non-overlapping for different $k$. This result implies that a projective measurement of the system observable can be realized through the apparatus pointer at orthogonality time.

With our results, we are able to provide a more detailed picture of how the measurement scheme of \cite{galapon1} works and confirm its attributes of inducing exact decoherence via one internal degree of freedom of the apparatus and exactly orthogonal pointer states. Note that these features of the scheme are realized provided that the probe and pointer are initially in momentum-limited states. One may ask if there are other initial conditions that can be imposed on the probe and pointer and therefore lead to exact decoherence. The Hamiltonian $H_M(t)$ given by equation (\ref{Hamiltonian}) is one of the possible measurement Hamiltonians that can be realized in the model of \cite{galapon1}. We may consider another measurement Hamiltonian given by  
\begin{equation}\label{Hamiltonian2} 
\tilde{H}_M(t) = \mu(t) [\gamma\, A \otimes P \otimes \mathbb{I}_{Po} + \delta\, \mathbb{I}_{\mathcal{S}} \otimes Q \otimes \Pi], 
\end{equation} 
where $\Pi$ is the momentum operator of the pointer, $\gamma$ and $\delta$ are positive coupling constants and $\mu(t)$ is a square pulse that is only non-zero within the time interval of interaction \cite{besagas}. The consequent quantum dynamics of $\tilde{H}_M(t)$ leads to exact decoherence when the probe is initially in a state $\langle p|\psi_{Pr}\rangle \in L^2(\mathbb{R})$ that is position-limited. Furthermore, the pointer states are exactly orthogonal provided a similar assumption on the initial state of the pointer. Our observations on the resulting quantum dynamics of $H_M(t)$ and $\tilde{H}_M(t)$ suggest the possibility that for every measurement Hamiltonian of a quantum measurement model, there corresponds an initial state or a set of conditions that, when imposed on the relevant subsystems, leads to exact decoherence and exactly orthogonal pointer states. While it is known that EIDT is being criticized for having features of only approximate suppression of coherences and approximately orthogonal pointer states, the results of \cite{galapon1} and this present work do not imply that EIDT is incorrect. In fact, the implications of these results may suggest an approach in coming up with EIDT models that exhibit exact decoherence. This can be done by further examination of the interaction Hamiltonian of every EIDT model and finding the corresponding initial state that must be imposed on the degrees of freedom of the environment in order for exact decoherence to occur and for the pointer states to become exactly orthogonal. 

We emphasize that coming up with a measurement model with decoherence being induced exactly (either via one or many degrees of freedom) only solves the problem of what process accounts for the exact quantum state collapse of the second kind. Indeed, the measurement model of \cite{galapon1} accomplishes the goal of solving of the problem of what accounts for the "exact" statistical collapse of the quantum state. However, this measurement model does not exhibit the quantum state collapse of the first kind. Therefore, the measurement problem still stands. In solving the entire measurement problem, we still have the remaining task of finding out a mechanism that accounts for the collapse of the first kind and perhaps by further investigation of the existing quantum mechanical interpretations.

\begin{appendices}

\section{Exact Particular Solution to Equations (\ref{vneqb}) and (\ref{vneqb2})}
The method of characteristics \cite{adzievski} is used to solve the first-order partial differential equations given by equations (\ref{vneqb}) and (\ref{vneqb2}). Rewriting equation (\ref{vneqb}) in the form
\begin{equation*}
\beta\,b\, g(t)\, \frac{\partial C_{kl}(q,b;t)}{\partial q} + \frac{\partial C_{kl}(q,b;t)}{\partial t} = -\frac{i}{\hbar} \alpha (a_k - a_l)\, g(t)\, q\, C_{kl}(q,b;t),
\end{equation*}
then the corresponding characteristic equation is
\begin{equation}\label{char}
\frac{\mathrm{d}q}{\beta b g(t)} = \frac{\mathrm{d}t}{1} = \frac{\mathrm{d}C_{kl}(q,b;t)}{-\frac{i}{\hbar} \alpha (a_k - a_l)\, g(t)\, q\, C_{kl}(q,b;t)}. \tag{A.1}
\end{equation} 
This is found to have linearly independent solutions
\begin{equation}
F(q, t, C_{kl}(q,b;t)) = q - \beta b g_0 t, \tag{A.2}
\end{equation}
and
\begin{equation}
G(q, t, C_{kl}(q,b;t)) = C_{kl}(q,b;t) \, \mathrm{e}^{\frac{i}{2\hbar}  \frac{\alpha}{\beta b} (a_k-a_l)q^2}. \tag{A.3}
\end{equation}
Then the general solution to equation (\ref{vneqb}) takes the form
\begin{equation}
C_{kl}(q,b;t) = u(q-\beta b g_0 t)\, \mathrm{e}^{-\frac{i}{2\hbar} \frac{\alpha}{\beta b} (a_k - a_l)q^2}, \tag{A.4}
\end{equation}
where $u(q-\beta b g_0 t)$ is a differentiable function of $q$ and $t$ whose explicit form can be obtained from the initial conditions. Applying the initial condition given by equation (\ref{inc}), then it can be shown that $u(q-\beta b g_0 t)$ is
\begin{multline}
u(q-\beta b g_0 t) =  \langle \varphi_k|\psi_{\mathcal{S}} \rangle \langle \psi_{\mathcal{S}} |\varphi_l \rangle \, |\langle q - \beta b g_0 t|\psi_{Pr}\rangle|^2 |\langle b|\Phi_{Po}\rangle|^2 \mathrm{e}^{\frac{i}{2\hbar} \frac{\alpha}{\beta b} (a_k - a_l)q^2}\,\\
\times \mathrm{e}^{-\frac{i}{\hbar} \alpha g_0 t (a_k - a_l)q}\,
\mathrm{e}^{\frac{i}{2\hbar} \alpha \beta g_0^2 t^2 (a_k - a_l)b}.\tag{A.5}
\end{multline}
Hence the particular solution to the von Neumann equation (equation \ref{vneqb}) is 
\begin{equation*}
C_{kl}(q,b;t) =  \langle \varphi_k|\psi_{\mathcal{S}} \rangle \langle \psi_{\mathcal{S}} |\varphi_l \rangle \, |\langle q - \beta b g_0 t|\psi_{Pr}\rangle|^2 |\langle b|\Phi_{Po}\rangle|^2 \, \mathrm{e}^{-\frac{i}{\hbar} \alpha g_0 t (a_k - a_l)q}\,
\mathrm{e}^{\frac{i}{2\hbar} \alpha \beta g_0^2 t^2 (a_k - a_l)b}.
\end{equation*}
Since, $C_{kl}(q,b;t) = f_k(q,b;t) f_l^*q,b;t)$, then
\begin{equation*}
f_k(q,b;t) = \langle \varphi_k|\psi_{\mathcal{S}} \rangle \langle q - \beta b g_0 t|\psi_{Pr}\rangle \langle b|\Phi_{Po}\rangle\, \mathrm{e}^{-\frac{i}{\hbar} \alpha g_0 t a_k q}\, \mathrm{e}^{\frac{i}{2\hbar} \alpha \beta g_0^2 t^2 a_k b}.
\end{equation*}

For equation (\ref{vneqb2}), the characteristic equation is
\begin{equation}
\frac{\mathrm{d}q}{\beta b g(t)} = \mathrm{d}t, \tag{A.6}
\end{equation}
which has a general solution
\begin{equation}
C_{kk}(q, b;t) = F(q -\beta b g_0 t).\tag{A.7}
\end{equation}
Applying the same initial condition (equation \ref{inc}) leads to the particular solution
\begin{equation*}
C_{kk}(q, b;t) = |\langle \varphi_k|\psi_{\mathcal{S}} \rangle|^2 \, |\langle q - \beta b g_0 t|\psi_{Pr}\rangle|^2\, |\langle b|\Phi_{Po}\rangle|^2
\end{equation*}

\section{Entire and Exponential Type Functions}
In this section we give a discussion of properties of entire and exponential type functions that are exploited in this work. Let $f(z)$ be a complex-valued function. This function is called entire and exponential of type $\tau > 0$ if and only if it satisfies the following properties: (i) $f(z)$ admits a series expansion $f(z) = \sum_{n=0}^{\infty} a_n z^n$ that is convergent in the entire complex plane, and (ii) for sufficiently large $|z|$, the inequality $|f(z)|<\mathrm{e}^{\tau|z|}$ holds \cite{boas, holland}. Given an entire function $f(z)$, the following theorem holds \cite{boas}:
\newtheorem{thm}{Theorem}[section]
\begin{thm}(Paley-Wiener)
	An entire function $f(z)$ is exponential of type $\tau > 0$ and belongs to $L^2$ in the real axis if and only if $f(z) = \int_{-\tau}^{\tau} \phi(t)\, \mathrm{e}^{itz} \mathrm{d}t$, where $\phi(t) \in L^2(-\tau, \tau)$. 
\end{thm}
The Paley-Wiener theorem has the following corollary: if a complex-valued function $f(z)$ is entire and exponential of type $\tau > 0$ and admits a real line restriction $f(x) \in L^2(\mathbb{R})$, then the complex-valued function $f(z+a)$ is entire and exponential of type $\tau > 0$ for every real $a$ \cite{galapon1}. 

In reference \cite{galapon1} and this paper, wave functions in the Hilbert space whose complex extensions are entire and exponential type are exploited. Here in particular, we consider the entire function
\begin{equation}\label{sinc}
\begin{split}
\psi(z) & = \frac{1}{\sqrt{\tau \pi}} \frac{\sin(\tau z)}{z},\\
		& = \frac{1}{2 \sqrt{\tau \pi}} \int_{-\tau}^{\tau} \mathrm{e}^{i k z}\, \mathrm{d}k,
\end{split}
\tag{B.1}
\end{equation}
for some $\tau > 0$. Then from the Paley-Wiener theorem, equation (\ref{sinc}) has a complex extension that is exponential of type $\tau$. Also, it can be shown that $\psi(z)$ satisfies $|\psi(z)| < \mathrm{e}^{\tau|z|}$ for sufficiently large $|z|$.

In Reference \cite{galapon1} and the current work, the products of entire and exponential type functions are also exploited. Let $f_1(x)$ and $f_2(x)$ be functions in the Hilbert space $L^2(\mathbb{R})$. Also, let the corresponding complex plane extensions $f_1(z)$ and $f_2(z)$, of $f_1(x)$ and $f_2(x)$, be entire and exponential with respective types $\tau_1$ and $\tau_2$. Then, the complex-valued function $f_1(z)f_2(z)$ is entire and exponential with type $\tau = \tau_1 + \tau_2$. Since, $f_1(z) = \int_{-\tau_1}^{\tau_1} \, \tilde{f}_1(t) \, \mathrm{e}^{itz}\, \mathrm{d}t$ and $f_2(z) = \int_{-\tau_2}^{\tau_2} \, \tilde{f}_2(t) \, \mathrm{e}^{itz}\, \mathrm{d}t$, then we can express  $f_1(z) f_2(z) = \int_{-(\tau_1 + \tau_2)}^{\tau_1 + \tau_2}\, \tilde{g}(u) \, \mathrm{e}^{izu}\, \mathrm{d}u$, where $\tilde{g}(u) = \frac{1}{2}\,\int_{-(\tau_1 + \tau_2)}^{\tau_1 + \tau_2}\, \tilde{f}_1((u+v)/2) \,\tilde{f}_2((u-v)/2)\, \mathrm{d}v$. With the use of Paley-Wiener theorem, then the product $f_1(z) f_2(z)$ is exponential with an overall type equal to $\tau_1 + \tau_2$ \cite{galapon1}.
\end{appendices}


\begin{thebibliography}{100}

\bibitem{zurek1}
W. H. Zurek, {\it Phys. Today}, {\bf 44}, 36 (1991)

\bibitem{zurek2}
W. H. Zurek, {\it Phys. Rev. D.} {\bf 24}, 4 (1991)

\bibitem{zurek3}
W. H. Zurek, {\it Phys. Rev. D.}, {\bf 26}, 8 (1992)

\bibitem{schlosshauer1}
M. Schlosshauer, {\it Rev. Mod. Phys.}, {\bf 76}, 1267 (2005)

\bibitem{schlosshauer2}
M. Schlosshauer, {\it Decoherence and the Quantum to Classical Transition} (Springer-Verlag, Berlin, Heidelberg) 2007.

\bibitem{galapon1}
E. A. Galapon, {\it EPL}, {\bf 113}, 60007 (2016)

\bibitem{clarke}
M. L. Clarke, {\it Eur. J. Phys.}, {\bf 35}, 015021 (2014) 

\bibitem{pessoa} O. J. Pessoa, {\it Synthese}, {\bf 113}, 323–346 (1998).

\bibitem{omnes}
R. Omn\`{e}s, {\it Found. Phys.}, {\bf 41}, 12 (2011)

\bibitem{namiki}
M. Namiki, {\it Found. Phys.}, {\bf 29}, 3 (1999)

\bibitem{wallace} D. Wallace, {\it Philosophical Transactions of the Royal Society A}, {\bf 370}, 4576-4593 (2012).

\bibitem{ballentine} L. Ballentine, {\it Foundations of Physics}, {\bf 38}, 916–922 (2008).

\bibitem{tanona} S. Tanona, {\it Synthese}, {\bf 190}, 3625-3649 (2013).

\bibitem{unruh and zurek}
W. Unruh, W.H. Zurek, {\it Phys. Rev. D}, {\bf 40}, 4 (1989)

\bibitem{venugopalan}
A. Venugopalan, D. Kumar, R. Ghosh, {\it Physica A}, {\bf 220}, 563-375 (1995)

\bibitem{connell}
R.F. O' Connell, J. Zuo, {\it Phys. Rev. A}, {\bf 67}, 062107 (2003)

\bibitem{ford}
G.W. Ford, R.F. O' Connell, {\it Am. J. Phys.}, {\bf 70}, 3 (2002)

\bibitem{myatt}
C. J. Myatt {\it et al.}, {\it Nature}, {\bf 403}, 269 (2000) 

\bibitem{abdul-redah}
T. Abdul-Redah, C.A. Chatzidimitriou-Dreismann, {\it App. Phys. A.}, {\bf 74}, S1379-S1381 (2002)

\bibitem{marques}
B. Marques {\it et al.}, {\it Scientific Reports} {\bf 5}, 16049 (2015)

\bibitem{arenz}
C. Arenz, R. Hillier, M. Fraas, D. Burgath, {\it Phys. Rev. A.}, {\bf 92}, 022102 (2015)

\bibitem{sun} 
C.P. Sun, D.L. Zhou, S.X.Yu, X.F.Liu, {\it Eur. Phys. J. D}, {\bf 13}, 145-155 (2001)

\bibitem{besagas}
J. P. A. Besagas {\it et al.}, in {\it Proceedings of the 36th Conference of Samahang Pisika ng Pilipinas} (Puerto Princesa City, Palawan, Philippines), SPP-2018-3B-02-1 

\bibitem{casas}
F. Casas, A. Nurua, M. Nadinic, {\it Comput. Phys. Commun.}, {\bf 183}, 2386 (2012)

\bibitem{busch}
P. Busch, P. Lahti, {\it Found. Phys.}, {\bf 26}, 875 (1996)

\bibitem{boas}
R.P.J. Boas, {\it Entire Functions}  (Academic Press Inc., Publishers, New York, NY, USA, 1954)

\bibitem{kofman}
A. Kofman, S. Ashab, F. Nori, {\it Phys. Rep.}, {\bf 520}, 2 (2012)

\bibitem{adzievski}
K. Adzievski, A.H. Siddiqi, {\it Introduction to Partial Differential Equations for Scientists and Engineers Using Mathematica}, (Chapman and Hall/CRC Press, Boca Raton, Florida, 2013)

\bibitem{griffiths}
D. Griffiths, {\it Introduction to Quantum Mechanics}, 2nd. ed., (Pearson, Upper Saddle River, NJ, 2005)

\bibitem{galapon2}
E. A. Galapon, arXiv:1709.08173, (2017)

\bibitem{lima}
J.C.L. Lima, {\it et al.} in {\it Proceedings of the 36th Conference of Samahang Pisika ng Pilipinas} (Puerto Princesa City, Palawan, Philippines), SPP-2018-PC-06-1 

\bibitem{holland}
A. Holland, {\it Introduction to the Theory of Entire Functions. Vol. 56}, (Academic Press, 1974)


\end{thebibliography}
\end{document}